\documentclass[useAMS, usenatbib]{article}
\usepackage{fullpage}
\usepackage{setspace}
\usepackage{comment}
\usepackage[pdftex]{graphicx}
\usepackage{amsmath}
\usepackage{amsthm}
\usepackage{amssymb}
\usepackage{graphicx}
\usepackage{fullpage}
\usepackage{setspace}
\usepackage{natbib}
\usepackage{color}

\usepackage{pgfplots}
\usepackage{tikz}
\usetikzlibrary{patterns}

\usepackage{algorithm}
\usepackage{algorithmic}

\newcommand{\rp}{\mathbb{R}^p}

\newcommand{\T}{\intercal}
\newcommand{\ideal}{\ensuremath{\mathrm{Ideal}}}

\renewcommand{\Re}{\ensuremath{\mathbb{R}}}

\DeclareMathOperator{\sgn}{\mathrm{sgn}}

\theoremstyle{definition}

\begin{document}

\title{Set-valued dynamic treatment regimes \\ for competing outcomes}

\author{Eric B. Laber\\Department of Statistics\\North Carolina State
University
  \and Daniel J. Lizotte \\Department of Computer Science \\ University of
Waterloo
  \and Bradley Ferguson\\ Department of Statistics \\ North Carolina
 State University}

\maketitle

\thispagestyle{empty}
\pagebreak

\begin{abstract}
  Dynamic treatment regimes operationalize the clinical decision
  process as a sequence of functions, one for each clinical decision,
  where each function takes as input up-to-date patient information
  and gives as output a single recommended treatment.  Current methods
  for estimating optimal dynamic treatment regimes, for example
  $Q$-learning, require the specification of a single outcome by which
  the `goodness' of competing dynamic treatment regimes are measured.
  However, this is an over-simplification of the goal of clinical
  decision making, which aims to balance several potentially competing
  outcomes. For example, often a balance must be struck between
  treatment effectiveness and side-effect burden.  We propose a method for
  constructing dynamic treatment regimes that accommodates competing
  outcomes by recommending sets of treatments at each decision
  point. Formally, we construct a sequence of set-valued functions
  that take as input up-to-date patient information and give as output
  a recommended subset of the possible treatments.  For a given patient history,
  the recommended set of treatments contains all treatments that are
  not inferior according to any of the competing outcomes.  When there is more than
  one decision point, constructing these set-valued functions requires
  solving a non-trivial enumeration problem.  We offer an exact
  enumeration algorithm by recasting the problem as a linear mixed
  integer program.  The proposed methods are illustrated using data
  from a depression study and from the CATIE schizophrenia study.
\end{abstract}
\thispagestyle{empty}
\pagebreak
\setcounter{page}{1}
\section{Introduction}
Dynamic treatment regimes (DTRs) attempt to operationalize the
clinical decision-making process wherein a clinician selects a treatment
based on current patient characteristics and then continues to adjust
treatment over time in response to the evolving heath status of the
patient.  Formally, a DTR is a sequence of decision rules, one for
each decision point, that take as input current patient information
and give as output a recommended treatment.  There is a growing
interest in estimating the DTRs from randomized or observational data,
typically with the goal of finding the DTR that maximizes the
expectation of a chosen clinical outcome.  A DTR is said to be optimal if
when followed by the population of interest it produces the maximal
clinical outcome on average.  Optimal DTRs have been estimated for the
management of a number of chronic conditions including ADHD
\citep{Laberetal, Nahum10, Nahum12}, depression 
\citep{schulte2012q, Song12}, 
HIV infection \citep{Demystifying},
schizophrenia \citep{shortreed11informing}, 
and cigarette addiction \citep{strecher2006moderators}.  
Approaches for estimating
optimal DTRs from data include $Q$-learning \citep{Nahum10, Watkins,
  WatkinsDayan}, $A$-learning \citep{Blatt, Murphy03}, 
regret regression \citep{Henderson},
and direct value maximization \citep{Orellana10, baqun, Zhao}.  

When estimating a decision rule from data using any of the
aforementioned methods, one must specify a single outcome and neglect
all others.  Perhaps the most obvious example is seeking the most
effective DTR without regard for side-effects.  Alternatively, one
might attempt to form a linear combination of two outcomes, e.g.\ side
effects and effectiveness, yielding a single composite outcome.
However, forming a composite outcome requires the elicitation of a
trade-off between two potentially incomparable outcomes.  For example,
one would need to know that a gain of 1 unit of effectiveness is worth
a cost of 3 units of side-effects.  Even if one could elicit this
trade-off at an aggregate level, assuming that a particular trade-off
holds for all future decision-makers is not reasonable since each will
have his or her own individual preferences which obviously cannot be
known \textit{a priori.}  \cite{lizotte12linear}
present one approach to dealing with this problem using a method that
estimates an optimal DTR for all possible linear trade-offs
simultaneously. Their method can also be used to explore what range of
trade-offs is consistent with each available treatment. Nonetheless,
their method must assume that there exists a \textit{linear} trade-off
that adequately describes any outcome preference, and they still
(perhaps implicitly) require the elicitation of this trade-off.

%Furthermore, since DTRs are typically
%estimated from data collected in an observational study or randomized
%trial with the intention of using the estimated DTR to inform the
%treatment of future patients, individual patient trade-offs are not
%available at the time the DTR is estimated.  

We propose an alternative approach for constructing DTRs that is
sensitive to competing outcomes but that avoids eliciting trade-offs
(linear or otherwise) between outcomes. Instead, our approach only
requires we elicit the size of a `clinically significant' difference
on each outcome scale.  Our proposed method still allows for the
incorporation of clinical judgement, individual patient preferences,
cost, and local availability, when no one treatment decision is best
across all competing outcomes.  Specifically, we propose set-valued
DTRs (SVDTRs) which, like DTRs, are a sequence of decision rules, one
for each time point.  However, the decision rules that comprise a
SVDTR take as input current patient information and give as output a
\textit{set} of recommended treatments.  The recommended treatment set
will be a singleton when there exists a treatment that is best across
all outcomes and will contain multiple treatments when no single
treatment is best.

The contributions of this paper are as follows. We introduce SVDTRs as
a new method for operationalizing sequential clinical decision making
that allows consideration of competing outcomes and the incorporation
of clinical judgment.  SVDTRs deal with competing outcomes without
having to elicit trade-offs between competing outcomes --- they only
require the elicitation of what constitutes a clinically significant
difference for each outcome individually.  We also provide a novel
mathematical programming formulation which allows us to efficiently
estimate SVDTRs from data.

The remainder of this paper is organized as follows.  In Section 2 we
review the $Q$-learning algorithm for estimating optimal DTRs from
data.  The $Q$-learning algorithm provides a starting point for the
construction of SVDTRs.  In Section~\ref{ss:singlepoint} we propose a
SVDTR for the single decision point problem. We then extend this
methodology to the two decision point problem in
Section~\ref{ss:twopoint}, and in Section~\ref{ss:computation} we
describe our mathematical programming approach for the efficient
estimation of SVDTRs from data.  In Section 5 we illustrate the
construction of SVDTRs using data from a single decision depression
trial \citep{keller2000comparison} and a two-stage schizophrenia study.
For clarity, the main body of the paper considers only binary
treatment decisions; we give an extension to an arbitrary number of
treatment options in the Appendix.

\section{Single outcome decision rules}\label{ss:singleoutcome}
In this section we review the $Q$-learning algorithm for estimating an
optimal DTR when there is a single outcome of interest.  For
simplicity, we will consider the case in which there are two decision
points and two treatment options at each decision point.  In this
setting the data available to estimate an optimal DTR are denoted by
$\mathcal{D} = \lbrace (H_{1i}, A_{1i}, H_{2i}, A_{2i},
Y_{i})\rbrace_{i=1}^{n}$ and consists of $n$ trajectories $(H_{1},
A_{1}, H_{2}, A_{2}, Y)$, one for each patient, drawn \textit{i.i.d.}\ from
some unknown distribution.  We use capital letters like $H_1$ and
$A_1$ to denote random variables and lower case letters like $h_1$ and
$a_1$ to denote realized values of these random variables.  The
components of each trajectory are as follows: $H_{t} \in
\mathbb{R}^{p_t}$ denotes patient information collected prior to the
assignment of the $t$th treatment, thus, this is information the
decision maker can use to inform the $t$th treatment decision; $A_t
\in \lbrace -1, 1\rbrace$ denotes the $t$th treatment assignment; $Y
\in \mathbb{R}$ denotes the outcome of interest which is assumed to be
coded so that higher values are more desirable than lower values.  The
outcome $Y$ is commonly some measure of treatment effectiveness but could
also be a composite measure attempting to balance different
objectives.

With a single outcome, the goal is to construct a pair of decision rules
$\pi = (\pi_1, \pi_2)$ where $\pi_t(h_t)$ denotes a decision rule for assigning 
treatment at time $t$ to a patient with history $h_t$ in such a way 
that the expected response $Y,$ given such treatment assignments, is maximized.
Formally, if $\mathbb{E}^{\pi}$ denotes the joint expectation over 
$H_t$, $A_t$, and $Y$ under the restriction that $A_t = \pi_t(H_t)$, then
the optimal decision rule $\pi^{\mathrm{opt}}$ satisfies 
$\mathbb{E}^{\pi^\mathrm{opt}}Y = \sup_{\pi}\mathbb{E}^{\pi}Y$.  Note that
the optimal decision rule defined in this way ignores the impact of 
the DTR $\pi^{\mathrm{opt}}$ on any other outcomes 
not incorporated into $Y$.  

One method for estimating an optimal DTR is the 
$Q$-learning algorithm \citep{Watkins, WatkinsDayan}.   
$Q$-learning is an approximate dynamic programming procedure that relies
on regression models to approximate the following conditional expectations 
\begin{eqnarray*}
Q_2(h_2, a_2) &\triangleq & \mathbb{E}(Y|H_2 = h_2, A_2=a_2), \\
Q_1(h_1, a_1) &\triangleq & \mathbb{E}\left(
\max_{a_2\in\lbrace -1,1\rbrace}Q_{2}(H_2, a_2)\big| H_1 = h_1, A_1 = a_1
\right).  
\end{eqnarray*}
The function $Q_t$ is termed the stage-$t$ $Q$-function.  The
function $Q_2(h_2, a_2)$ measures the quality of assigning treatment
$a_2$ at the second decision point to a patient with history $h_2$.
The function $Q_1(h_1, a_1)$ measures the quality of assigning
treatment $a_1$ at the first decision point to a patient with history
$h_1$ assuming that an optimal treatment decision will be made at the
second decision point.  From these definitions it is clear that
$\pi_2^{\mathrm{opt}}(h_2) = \arg\max_{a_2\in\lbrace -1,1\rbrace}Q_2(h_2,a_2)$, and,
assuming that $\pi_2^{\mathrm{opt}}$ is followed at the second decision
point, $\pi_1^{\mathrm{opt}}(h_1) = \arg\max_{a_1\in\lbrace -1,1\rbrace}Q_1(h_1,a_1)$.
Note that this is nothing more than the dynamic programming solution to 
finding the optimal sequence of decision rules \citep{bellman}.  

In practice, the $Q$-functions are not known and so a natural approach
is to estimate them from data.  As is common in practice, we will
consider linear working models of the form 
$Q_t(h_t, a_t) = h_{t,1}^{\T}\beta_{t} + a_th_{t,2}^{\T}\psi_{t}$,
where $h_{t,1}$ and $h_{t,2}$ are (possibly the same) subvectors of $h_{t}$.  
The $Q$-learning algorithm proceeds in three steps:
\begin{enumerate}
\item Estimate the parameters indexing the working model for the
  stage-2 $Q$-function using least squares.  Let $\hat{\beta}_{2}$ and
  $\hat{\psi}_{2}$ denote the corresponding estimators, and let
  $\hat{Q}_2(h_2, a_2)$ denote the fitted model.  
\item 
  \begin{enumerate}
    \item Define the predicted future outcome $\tilde{Y}$ following the
      estimated optimal decision rule at stage two as
      $\tilde{Y} \triangleq \max_{a_2\in\lbrace -1,1\rbrace}\hat{Q}_{2}(H_2, a_2)$.
    \item Estimate the parameters indexing the working model for the
      stage-1 $Q$-function using least squares.  That is, regress
      $\tilde{Y}$ on $H_1$ and $A_1$ using the working model to obtain
      $\hat{\beta}_{1}$ and $\hat{\psi}_{2}$.  Let $\hat{Q}_1(h_1, a_2)$ denote
      the fitted model.  
  \end{enumerate}
\item Define the $Q$-learning estimated optimal treatment regime 
  $\hat{\pi} = (\hat{\pi}_1, \hat{\pi}_2)$ so that
  $\hat{\pi}_{t}(h_t) = \arg\max_{a_t\in\lbrace -1,1\rbrace}\hat{Q}(h_t, a_t)$.  
\end{enumerate}
The $Q$-learning algorithm is simple to implement and easy to
interpret given its connections to dynamic programming.  For these
reasons we use $Q$-learning as the basis for developing SVDTRs 
which we introduce in the next section. 
However, $Q$-learning is not the only procedure for estimating an 
optimal DTR.  Alternatives to $Q$-learning 
include $A$-learning \citep{Blatt, Murphy03, schulte2012q},
regret regression \citep{Henderson}, and penalized $Q$-learning
\citep{Song12}.  

Penalized $Q$-learning also lends itself
to producing SVDTRs, though of a different 
nature.  Briefly, penalized $Q$-learning employs an unusual singular 
penalty to estimate the coefficients in the second stage $Q$-function. 
For a given tuning parameter $\lambda > 0$ estimated coefficients
$\tilde{\beta}_2$ and $\tilde{\psi}_2$ satisfy
\begin{equation*}
(\tilde{\beta}_2^{\T}, \tilde{\psi}_{2}^{\T})^{\T} = \arg\min_{\beta_2, \psi_2\in
\mathbb{R}^{p_2}} \sum_{i=1}^{n}(Y_{i} - H_{2,1,i}^{\T}\beta_2 - 
A_{2i}H_{2,2,i}^{\T}
\psi_2)^2 + \lambda\sum_{i=1}^{n}|H_{2,2,i}^{\T}\psi_{2}|.
\end{equation*}
Using this approach, under certain generative models, 
$H_{2,2}^{\T}\tilde{\psi}_{2}$ will be exactly zero for a non-null set of 
$H_{2}$ values \citep[see][for details]{Song12}.
One can then define the 
set-valued second stage decision rule 
\begin{equation*}
\tilde{\pi}(h_2) = \left\lbrace \begin{array}{c}
\lbrace \sgn(h_{2,2}^{\T}\tilde{\psi}_{2})\rbrace,\,\,\mathrm{if}\,\,
h_{2,2}^{\T}\tilde{\psi}_{2} \ne 0, \\ 
\lbrace -1, 1 \rbrace,\,\,\mathrm{otherwise.} 
\end{array}
\right.
\end{equation*}
The foregoing set-valued decision rule assigns a single treatment for
second stage histories $h_2$ that, based on the estimated coefficient
$\tilde{\psi}_{2}$ have a nonzero treatment effect.  On the other
hand, if the estimated treatment effect $h_{2,2}^{\T}\tilde{\psi}_2$ is
zero, then both treatments are recommended.  An analogous approach could
be used to form a set-valued decision rule at the first stage.  

In the preceding development we have adapted the ideas of
~\cite{Song12} to suit our purposes.  They proposed penalized $Q$-learning
in an effort to improve coverage probabilities of confidence intervals
for first stage coefficients.  Thus, any problems with this
development are ours and should not be attributed to 
~\cite{Song12}.  Secondly, such set-valued treatment regimes attempt to
recommend sets of treatment when there is insufficient evidence of a
significant treatment effect with respect to a single outcome measure
for a patient with a given history.  This should be contrasted with
our goal of balancing the treatment effects of competing outcomes.

%\begin{itemize}
%  \item Introduce and discuss $Q$-learning algorithm 
%  \item Reference work by Song et al. (2011) as a method of
%    potentially constructing a set-valued dynamic treatment regime
%    where the set is identified as patients with no treatment effect
%    for the specified univariate outcome.  Also discuss the obvious
%    hypothesis testing approach to this problem. Make sure we're not 
%    stepping on Tianshuang's work---we must be very clear that what
%    we're attacking a fundamentally different problem.  
%  \item Note: this section needs to be terse since we won't introduce 
%    our method until section 3 which is quite late. 
%\end{itemize}

\section{Static set-valued decision rules}\label{ss:singlepoint}
In this section we discuss the estimation of a set-valued decision
rule when there is a single decision point and there are two competing
outcomes of interest (see the Appendix for the generalization to an
arbitrary number of treatment options).  The data available to
estimate the decision rule is denoted by $\mathcal{D} = \lbrace (H_i,
A_i, Y_i, Z_i)\rbrace_{i=1}^n$ and is comprised of $n$ trajectories
$(H, A, Y, Z)$ drawn independently from the same distribution.  The
elements of each trajectory are as follows: $H$ denotes the
information available to the decision maker \textit{before} the
assignment of treatment and is assumed to take values in $\rp$; $A$
denotes the randomly assigned treatment which is assumed to be binary
and coded to take values in the set $\lbrace -1, 1\rbrace$; $Y$
denotes the first outcome of interest which is assumed to take values
in $\mathbb{R}$ and is coded so that higher values of $Y$ correspond
to more desirable clinical outcomes; and $Z$ denotes the second
outcome of interest which is assumed to take values in $\mathbb{R}$
and is also coded so that higher values are more desirable.  It is
also assumed that one has obtained, either by elicitation or
historical data, the positive quantities $\Delta_{Y}$ and $\Delta_{Z}$
denoting `clinically meaningful differences' in the outcomes $Y$ and
$Z$ respectively.  That is, a clinician would be willing to change a
patient's current treatment if this change yielded a difference of at
least $\Delta_Y$ ($\Delta_Z$) in the outcome $Y$ ($Z$) and all other
things were held equal.  Note that in eliciting $\Delta_Y$ there is no
need to reference the competing outcome $Z$ and vice versa when
eliciting $\Delta_{Z}$.

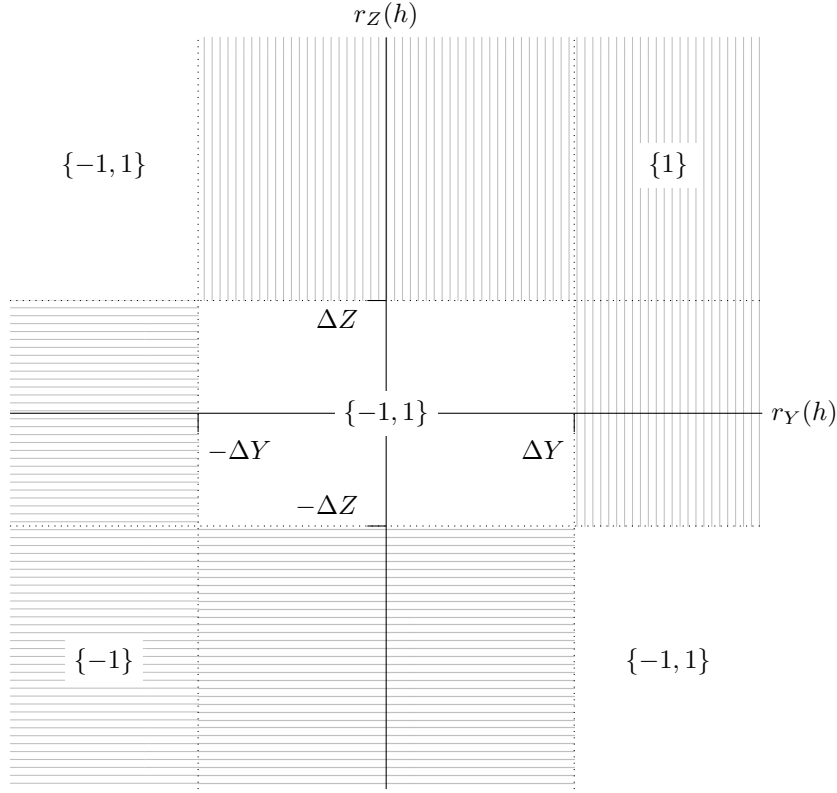
\begin{figure}
\usetikzlibrary{patterns}
\begin{center}
\begin{tikzpicture}[baseline,scale=5]

%Central region of indifference
%\pattern[pattern=dots,pattern color=lightgray] (-0.5,-0.3) rectangle (0.5,0.3);
%Lower-right
%\pattern[pattern=dots,pattern color=lightgray] (0.5,-0.3) rectangle (1,-1);
%Upper-left
%\pattern[pattern=dots,pattern color=lightgray] (-0.5,0.3) rectangle (-1,1);

%Prefer treatment 1
\pattern[pattern=vertical lines,pattern color=lightgray] (-0.5,0.3) rectangle (0.5,1);
\pattern[pattern=vertical lines,pattern color=lightgray] (0.5,0.3) rectangle (1,1);
\pattern[pattern=vertical lines,pattern color=lightgray] (0.5,-0.3) rectangle (1,0.3);

%Prefer treatment -1
\pattern[pattern=horizontal lines,pattern color=lightgray] (-0.5,0.3) rectangle (-1,-0.3);
\pattern[pattern=horizontal lines,pattern color=lightgray] (-0.5,-0.3) rectangle (-1,-1);
\pattern[pattern=horizontal lines,pattern color=lightgray] (0.5,-0.3) rectangle (-0.5,-1);

%Y direction (horizontal)
\draw (-1,0) -- (1,0);

%Delta Z bounds
\draw[dotted] (-1,0.3) -- (1,0.3);
\draw[dotted] (-1,-0.3) -- (1,-0.3);

%Z direction (vertical)
\draw (0,-1) -- (0,1);

%Delta Y bounds
\draw[dotted] (0.5,-1) -- (0.5,1);
\draw[dotted] (-0.5,-1) -- (-0.5,1);

%Delta Y ticks
\draw (0.5,0) -- (0.5,-0.05) node[anchor=north east] {$\Delta Y$};
\draw (-0.5,0) -- (-0.5,-0.05) node[anchor=north west] {$-\Delta Y$};

%Delta Z ticks
\draw (0,0.3) -- (-0.05,0.3) node[anchor=north east] {$\Delta Z$};
\draw (0,-0.3) -- (-0.05,-0.3) node[anchor=south east] {$-\Delta Z$};

%Treatment labels
\node at (-0.75,0.66) [fill=white] {$\{-1,1\}$};
\node at (0,0) [fill=white] {$\{-1,1\}$};
\node at (0.75,-0.66) [fill=white] {$\{-1,1\}$};

%Axis labels
\node at (0,1) [anchor=south] {$r_Z(h)$};
\node at (1,0) [anchor=west] {$r_Y(h)$};

\node at (-0.75,-0.66) [fill=white] {$\{-1\}$};
\node at (0.75,0.66) [fill=white] {$\{1\}$};
\end{tikzpicture}
\end{center}
\caption{\label{fig:rtopi}Diagram showing how the output of $\pi^\ideal(h)$ depends on
  $\Delta_Y$ and $\Delta_Z$, and on the location of the point $(r_Y(h),r_Z(h))$.}
\end{figure}

The goal is to construct a decision rule $\pi:\rp \rightarrow \lbrace
\lbrace -1 \rbrace, \lbrace 1 \rbrace, \lbrace -1, 1 \rbrace \rbrace$
that maps baseline patient information $H$ into a subset of the
available treatment decisions.  Ideally, for a given baseline history
$h$, the decision rule $\pi$ would recommend a single treatment if
that treatment was expected to yield a clinically meaningful
improvement in at least one of the outcomes and, in addition, that
treatment was not expected to lead to a significant detriment in terms
of the other outcome.  On the other hand, if this cannot be said of
one the treatments then the decision rule should instead return the set
$\lbrace -1, 1\rbrace$ and leave the `tie-breaking' to the decision
maker.  Put more formally, if we define the (non-normalized) treatment
effects for each outcome as $r_Y(h) \triangleq \mathbb{E}(Y|H=h, A=1)
- \mathbb{E}(Y|H=h, A=-1)$ and similarly $r_Z(h) \triangleq
\mathbb{E}(Y|H=h, A=1) - \mathbb{E}(Y|H=h, A=-1)$, then the ideal
decision rule satisfies
\begin{equation}\label{piOneStageIdeal}
\pi_{\Delta}^{\ideal}(h) = \left\lbrace
\begin{array}{l}
\lbrace\sgn(r_Y(h))\rbrace,\,\,\mathrm{if}\,\, |r_Y(h)| \ge \Delta_Y\,\,
\mathrm{and}\,\,
\sgn(r_Y(h))r_Z(h) > -\Delta_Z, \\ 
\lbrace\sgn(r_Z(h))\rbrace,\,\,\mathrm{if}\,\, |r_Z(h)| \ge \Delta_Z\,\,
\mathrm{and}\,\,
\sgn(r_Z(h))r_Y(h) > -\Delta_Y, \\
\lbrace -1, 1\rbrace, \,\,\mathrm{otherwise,}\,\,
\end{array}
\right.
\end{equation}
where $\sgn$ denotes the signum
function.  Figure~\ref{fig:rtopi} illustrates how $\pi_{\Delta}^{\ideal}(h)$
depends on $r_Y(h)$, $r_Z(h)$, $\Delta_Y$, and $\Delta_Z$. If we
consider the point $(r_Y(h),r_Z(h)) \in \Re^2$, its location relative
to the points
$(\Delta_Y,\Delta_Z)$, $(-\Delta_Y,\Delta_Z)$, $(\Delta_Y,-\Delta_Z)$
and $(-\Delta_Y,-\Delta_Z)$ determines whether we prefer treatment $1$,
prefer treatment $-1$, or are indifferent according to the criteria set
out above.

Following the
motivation for $Q$-learning, we will estimate the ideal decision by
modelling the conditional expectations $Q_Y(h, a) \triangleq
\mathbb{E}(Y|H=h, A=a)$ and $Q_Z(h,a) \triangleq \mathbb{E}(Z|H=h,
A=a)$.  We will use linear working models of the form
\begin{eqnarray}\label{workingQModelsSetValuedI}
Q_Y(h, a) &=& h_{1,1,Y}^{\T}\beta_{Y} + ah_{1,2,Y}^{\T}\psi_{Y}, \\ 
\label{workingQModelsSetValuedII}
Q_Z(h, a) &=& h_{1,1,Z}^{\T}\beta_{Z} + ah_{1,2,Z}^{\T}\psi_{Z}, 
\end{eqnarray}
where $h_{1,1,Y}$, $h_{1,2,Y}$, $h_{1,1,Z}$, and $h_{1,2,Z}$ are
(possibly the same) subvectors of $h$, and estimate the coefficients
$\beta_Y, \psi_Y, \beta_Z$, and $\psi_Z$ using least
squares.   In the remainder of this Section,  
in an effort to avoid cumbersome notation, we will assume
$h_{i,j,Z}$ and $h_{i,j,Y}$ are equal and thus can be denoted as $h_{i,j}$.
This assumption is only for notational efficiency and is not used
in Section 5.     
Let $\hat{\beta}_{Y}, \hat{\psi}_{Y}, \hat{\beta}_{Z}$, and
$\hat{\psi}_{Z}$ denote the corresponding least squares estimators.
Note that the implied estimators of $r_{Y}(h)$ and $r_{Z}(h)$ are
$2h_{2,1}^{\T}\hat{\psi}_Y$ and $2h_{2,1}^{\T}\hat{\psi}_Z$ respectively.  Hence,
a simple plug-in estimator of the ideal decision rule is given by
\begin{equation}\label{staticOLSOpt}
\hat{\pi}_{\Delta}(h) = \left\lbrace
\begin{array}{l}
\lbrace \sgn(h_{1,2}^{\T}\hat{\psi}_{Y})\rbrace ,\,\,
\mathrm{if}\,\, 2|h_{1,2}^{\T}\hat{\psi}_{Y}| \ge \Delta_Y\,\, \mathrm{and} \,\,
\sgn(h_{1,2}^{\T}\hat{\psi}_{Y})2h_{1,2}^{\T}\hat{\psi}_{Z} > -\Delta_Z, \\
\lbrace \sgn(h_{1,2}^{\T}\hat{\psi}_{Z})\rbrace ,\,\,
\mathrm{if}\,\, 2|h_{1,2}^{\T}\hat{\psi}_{Z}| \ge \Delta_{Z}\,\, 
\mathrm{and} \,\, \sgn(h_{1,2}^{\T}\hat{\psi}_{Z})2h_{1,2}^{\T}\hat{\psi}_{Y} 
> -\Delta_Y, \\
\lbrace -1, 1\rbrace , \,\, \mathrm{otherwise}. 
\end{array}
\right.
\end{equation}
The above decision rule mimics the ideal decision rule and is easily shown
to be consistent under mild moment conditions and the assumption 
that the working models in (\ref{workingQModelsSetValuedI}) and 
(\ref{workingQModelsSetValuedII}) are correct.  While the above
model is a compact representation of $\hat{\pi}_{\Delta}$, it will
often be convenient to display the decision rule either as a tree 
or as regions in the $h_{1,2}^{\T}\hat{\psi}_{Y},\, h_{1,2}^{\T}\hat{\psi}_{Z}$ plane
(see Section 5 for examples).

\subsection{Preference heterogeneity and set-valued rules} 
Personalized medicine recognizes the need to account for heterogeneity
in treatment effects across patients.   
However, it is important to recognize not only that
different patients will experience different outcomes under the same treatments, 
but also that different patients will rate \textit{the same outcomes} 
differently.  In this section we illustrate how set-valued decision rules can
accommodate patient individual preference even though such preferences 
cannot be known at the time the data are collected and the models 
are estimated.  We contrast set-valued decision rules with single outcome
decision rules formed using a composite outcome.  We shall see that
when the composite 
outcome closely reflects patient preferences then the corresponding 
decision rule performs well, however, when the composite outcome does
not reflect patient preferences, then the quality of the single outcome
decision rule can perform poorly.

We will consider data generated from the following simple class of 
generative models
\begin{equation*}
\begin{array}{l}
H = (H_1, H_2)^{\T} \sim \mathrm{Normal}_{2}
\left(
0, \Omega
\right), \\ A \sim \mathrm{Uniform}\lbrace-1, 1\rbrace,  \\ 
Y = A(\psi_{Y,1} + \psi_{Y,2}H_1 + \psi_{Y,3}H_2 + \psi_{Y,4}H_1H_2), \\ 
Z = A(\psi_{Z,1} + \psi_{Z,2}H_1 + \psi_{Z,3}H_2 + \psi_{Z,4}H_1H_2),
\end{array}
\end{equation*}
where $\Omega_{1,1}=\Omega_{2,2} = 1$, and
$\Omega_{1,2}=\Omega_{1,2} = \rho$. 
Thus, the class of models
is determined by $\psi_{Y}$, $\psi_{Z}$, $\rho$, and the thresholds 
$\Delta_{Y}$ and $\Delta_{Z}$.  
Note that we have omitted
specifying a main-effect term in the conditional mean of the outcomes
$Y$ and $Z$ since these do not affect the optimal decision rule; similarly,
we have assumed that $Y$ and $Z$ are observed without error since an 
independent additive error would not affect the optimal decision rule. 
We vary the parameters indexing the generative models in order
to highlight factors that affect the performance of the set-valued
decision rule.  In particular, we systematically vary the following
three components of the generative model: (i) the proportion of 
individuals for which there is a unique best treatment option (\texttt{Uniq}); 
(ii) the
proportion of individuals for which neither treatment yields a significant
treatment effect on either outcome (\texttt{Null}); and (iii) the proportion of 
individuals for which there are significant treatment effects for 
both outcomes but the effects run in opposite directions (\texttt{Opst}).  More
specifically, define
\begin{eqnarray*}
\mathrm{\texttt{Uniq}} &\triangleq &
P\bigg(\big\lbrace
|r_{Y}(H)| \ge \Delta_{Y}, \mathrm{sgn}(r_{Y}(H))r_{Z}(H) \ge -\Delta_{Z}
\big\rbrace \bigcup \\ && \quad \quad \quad
\big\lbrace
|r_{Z}(H)| \ge \Delta_{Z}, \mathrm{sgn}(r_{Z}(H))r_{Y}(H) \ge -\Delta_{Y})
\big\rbrace
\bigg); \\
%\end{multline*}
%\begin{equation*}
\mathrm{\texttt{Null}} &\triangleq& P\left(
\lbrace |r_{Y}(H)| < \Delta_{Y}\rbrace \bigcap
\lbrace |r_{Z}(H)| < \Delta_{Z}\rbrace
\right); \, \mathrm{and}\\
\mathrm{\texttt{Opst}} &\triangleq &
P\left(
|r_{Y}(H)| \ge \Delta_{Y},\, |r_{Z}(H)| \ge \Delta_{Z},\,
r_{Y}(H)r_{Z}(H) < 0
\right).
\end{eqnarray*}
Note that \texttt{Uniq}, \texttt{Null}, \texttt{Opst} sum to one.  
The three settings for \texttt{Uniq}, \texttt{Null}, and \texttt{Opst} that
we consider here and the corresponding values of $\psi_{Y}, \psi_{Z}, \rho, 
\Delta_{Y}$, and $\Delta_{Z}$ are given in Table \ref{pseudoSimTable}.
%Results for additional parameter settings can be found in the appendix.  

\begin{table}[here]\label{pseudoSimTable}
  \begin{center}
  \begin{tabular}{ccccccccc}
    Setting  & 
    \texttt{Uniq} & \texttt{Null} & \texttt{Opst} & $\psi_{Y}$ & $\psi_{Z}$
    & $\rho$ & $\Delta_{Y}$ & $\Delta_{Z}$ \\ \hline 
    1 & $0.80$ & $0.10$ & $0.10$ & $(-0.30, 0.25, -2.0)$ & 
    $(0.0, -0.72, -0.74)$ & $-0.38$ & $0.5$ & $0.5$ \\  
    2 & $0.45$ & $0.10$ & $0.45$ & $(-0.05, 0.40, -1.25)$ & 
    $(0.65, -0.85, 0.29)$ & $-.36$ & $0.5$ & $0.5$ \\ 
    3 & $0.10$ & $0.10$ & $0.80$ & $(-1.0, -1.4, 2.0)$ & 
    $(1.6, 2.2, -2.2)$ & $-0.4$ & $0.5$ & $1.0$ 
  \end{tabular}
  \end{center}
  \caption{Settings for parameters indexing the underlying generative
    model.  The settings vary from mostly individuals with
    unique best treatments to 
    mostly individuals with significant treatment effects running in opposite 
    directions.} 
\end{table}

In this example, patient preference is operationalized
through what we term the preference parameter $\delta \in [0,1]$,
similar to the approach of \cite{lizotte12linear}.  A
preference of $\delta$ indicates that a patient would be ambivalent
between a one unit improvement (detriment) in $Z$ and a
$\delta/(1-\delta)$ improvement (detriment) in $Y$.  For patients 
with preference parameter $\delta$ the optimal decision rule is 
given by $\pi_{c,\delta}(h) \triangleq 
\arg\max_{a}\mathbb{E}(\delta Y + (1-\delta)Z | H=h,
A=a)$; thus, the optimal composite outcome is given by 
$\delta Y + (1-\delta)Z$.  Below we will compare decision rules 
derived from the set-valued decision rule of the preceding section with  
$\pi_{c, 0.5}$ and $\pi_{c, 0.25}$.

For a patient's `true preference' $\delta^*$ define
the \textit{regret} of an arbitrary decision rule $\pi$ as 
\begin{equation*}
\mathbb{E}\left[
\mathbb{E}(\delta^*Y + (1-\delta^*)Z|H, A=\pi(H))
-\max_{a}\mathbb{E}(\delta^*Y + (1-\delta^*)Z|H, A=a)
\right],
\end{equation*}
so that the regret measures the average loss in performance incurred by
applying a suboptimal decision rule $\pi$.  The regret is nonnegative
and equals zero when $\pi$ agrees with the optimal decision rule,
$\pi_{c,\delta^*}$, almost surely.  

In order to define the regret for a set-valued decision rule one must
specify a mechanism for choosing a treatment from the set of
recommended treatments.  Here we consider two possible `tie-breaking'
scenarios.  In the first, we assume that the clinician will choose the
best action from among the recommended treatments with probability
0.75; we believe this reflects a clinicians ability to leverage
individual patient characteristics and preferences in the decision
process.  Recall that the set-valued decision rule provides not only
the pool of recommended treatments but also the estimated mean
outcomes for each response; thus this information can be used to
inform the clinician's decision.  We term the resultant (random)
decision rule the 75\% optimal compatible policy.  In the second
tie-breaking scenario we imagine an adversarial decision maker that
always chooses the worst of the available treatments.  Such a decision
maker was considered by \cite{fard11nondeterministic} in the study of
non-deterministic decision rules for a single outcome.  While a
clinician that is actively working against their patients is
unrealistic, the performance of such a policy is useful for
illustrating the impact of screening out suboptimal treatments which
occurs in the formation of the set-valued decision rule.  We term the
resultant (deterministic) decision rule the 0\% optimal compatible
policy.  To provide additional baselines for comparison with the 75\%
and 0\% optimal compatible polices, we also consider a policy in which
a clinician chooses the optimal treatment from among \textit{all}
possible treatments 75\% of the time and a policy in which the
clinician always chooses the worst possible treatment from among
\textit{all} possible treatments; we term these policies the 75\%
optimal policy and 0\% optimal policy respectively.

\begin{figure}
\begin{minipage}{.28\linewidth}
\includegraphics[width=2.5in]{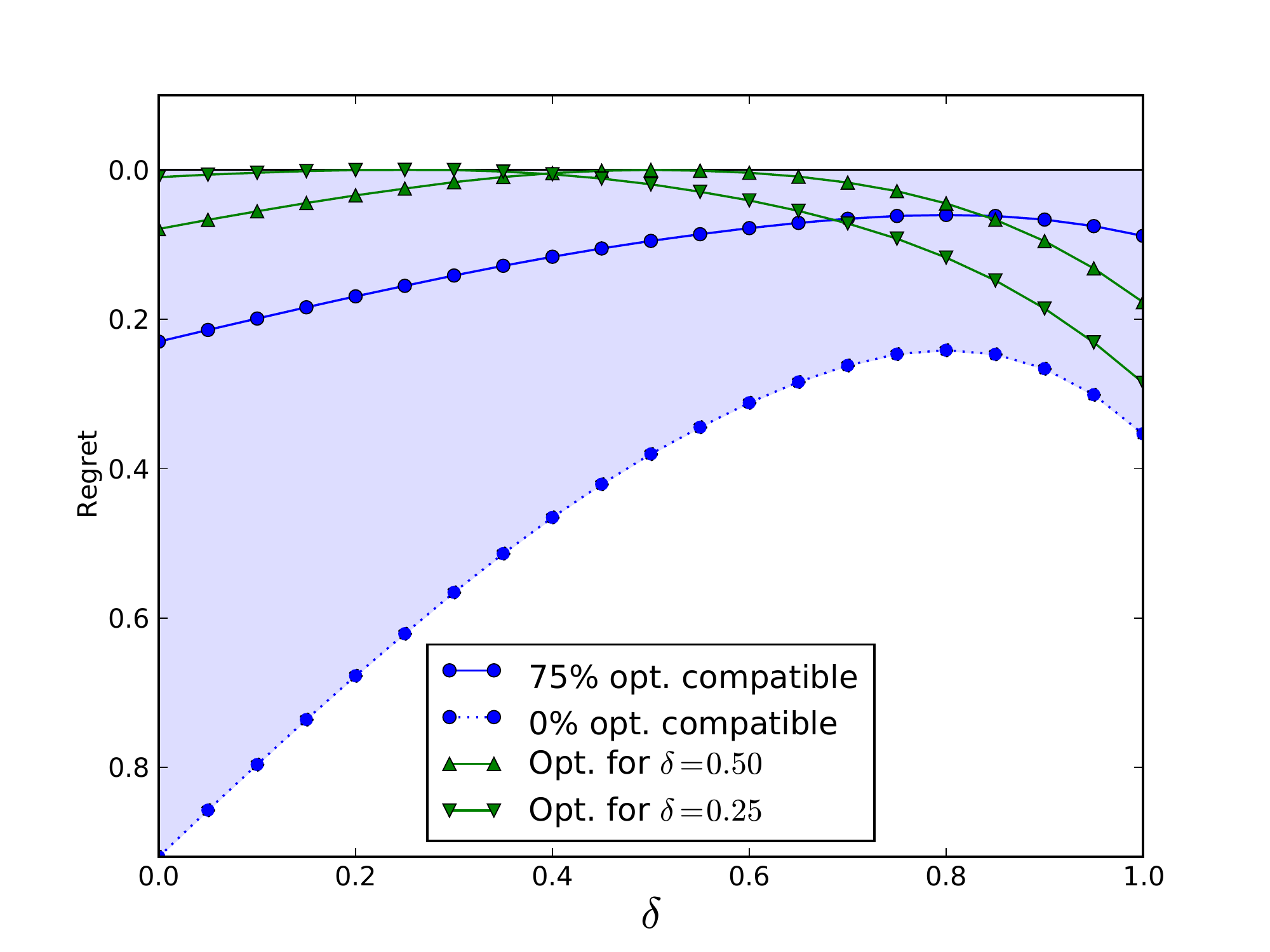}
\end{minipage}
\hspace{.05\linewidth}
\begin{minipage}{.28\linewidth}
\includegraphics[width=2.5in]{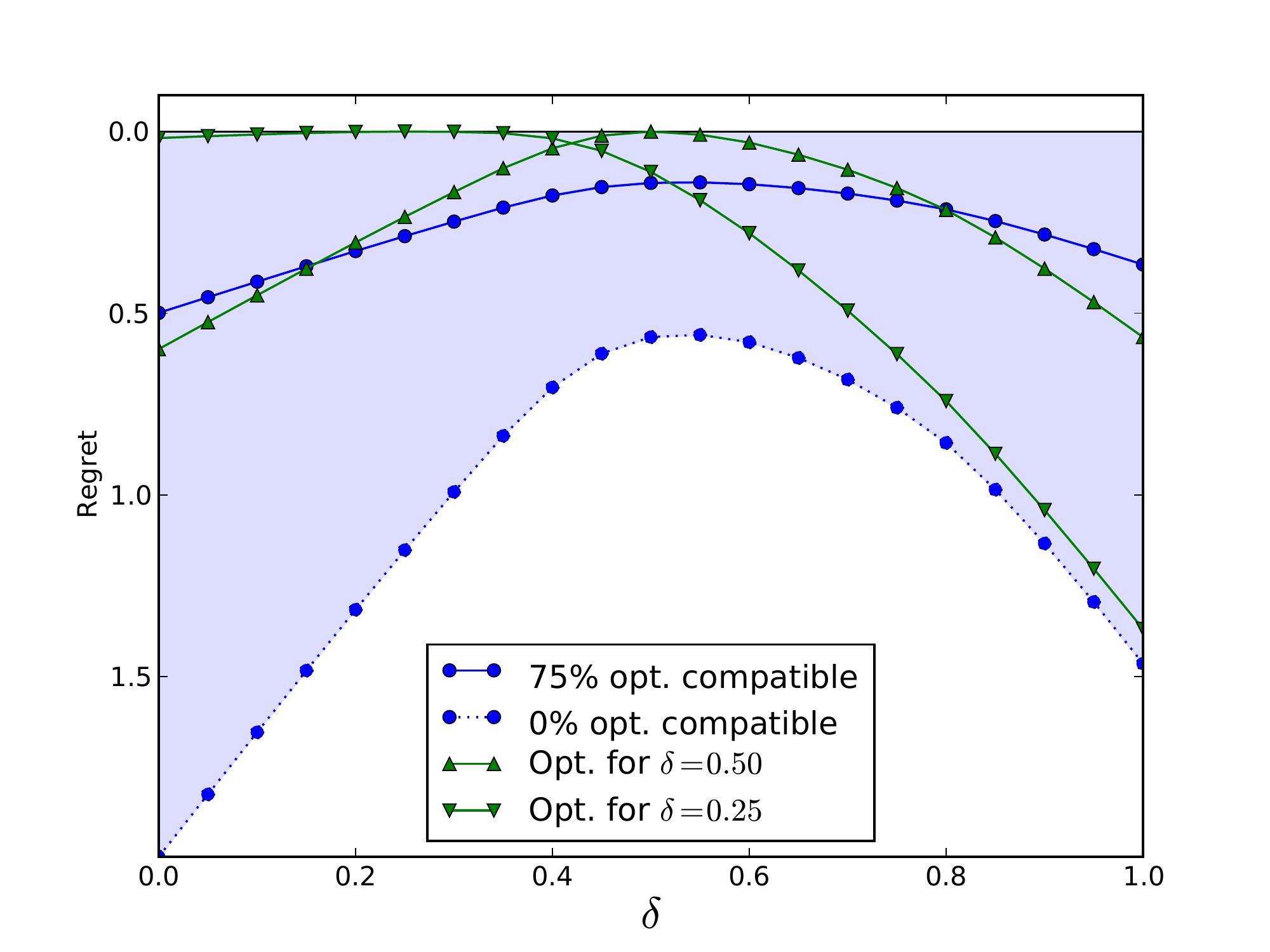}
\end{minipage}
\hspace{.05\linewidth}
\begin{minipage}{.28\linewidth}
\includegraphics[width=2.5in]{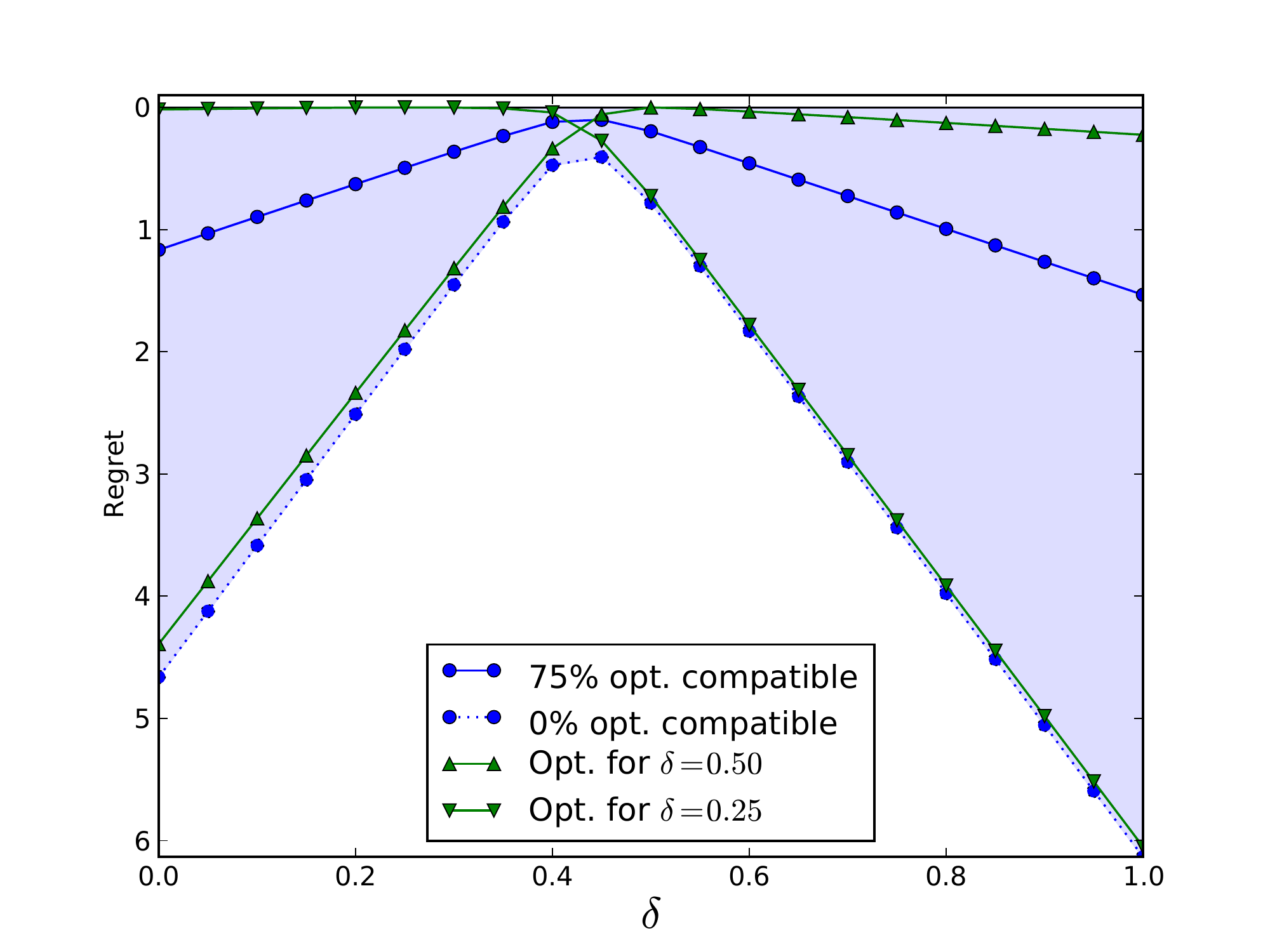}
\end{minipage}
\label{greenVBluePlots}
\caption{Regret versus patient preference for the 75\%
  and 0\% consistent policies as well as $\pi_{c, 0.5}$ and $\pi_{c,
    0.25}$.  The blue shaded region indicates the space of all 
  policies that are compatible with the set-valued decision rule.  
  From left to right: parameter settings 1, 2, and 3.}  
\end{figure}

\begin{figure}
\begin{minipage}{.28\linewidth}
\includegraphics[width=2.5in]{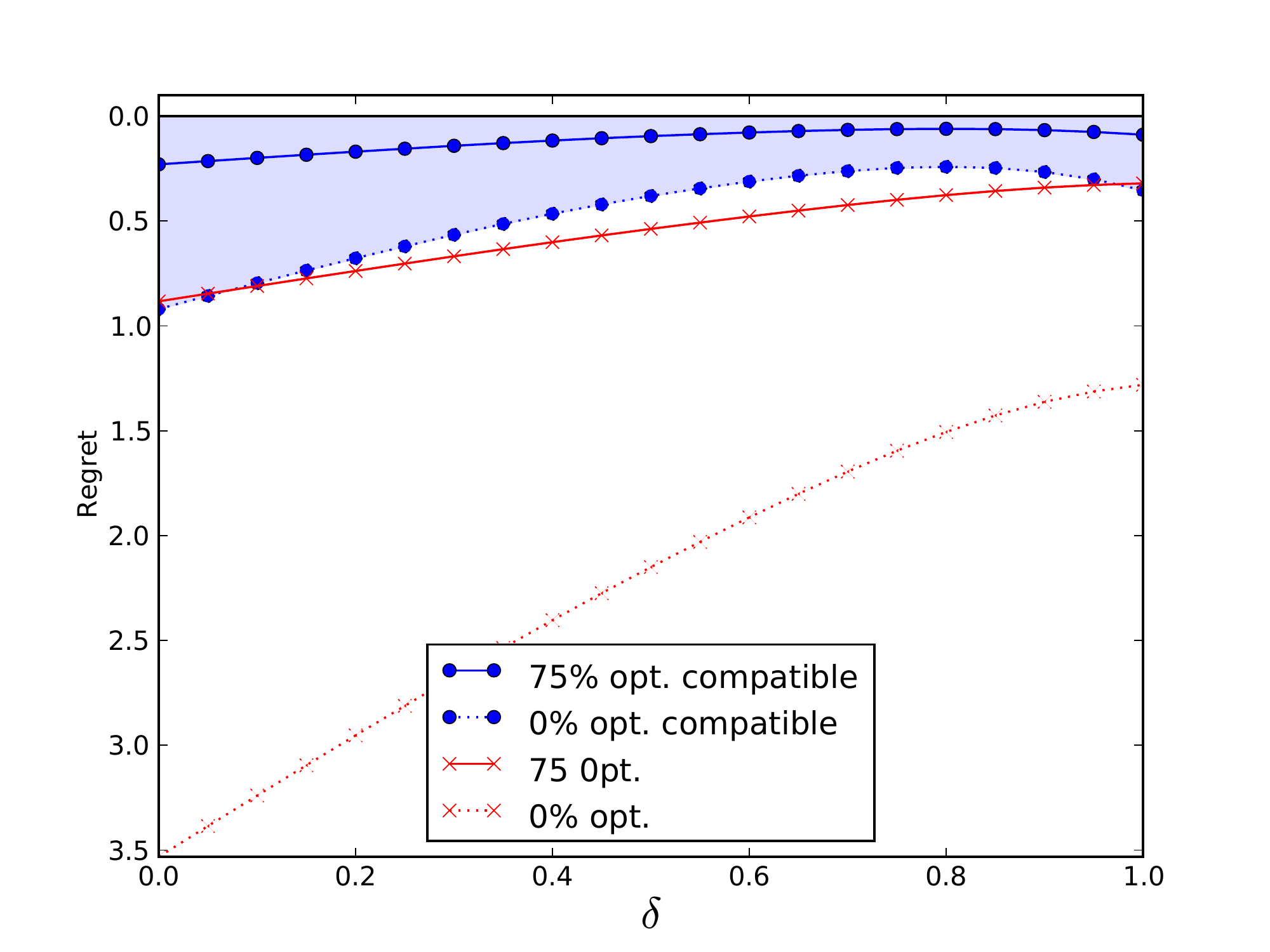}
\end{minipage}
\hspace{.05\linewidth}
\begin{minipage}{.28\linewidth}
\includegraphics[width=2.5in]{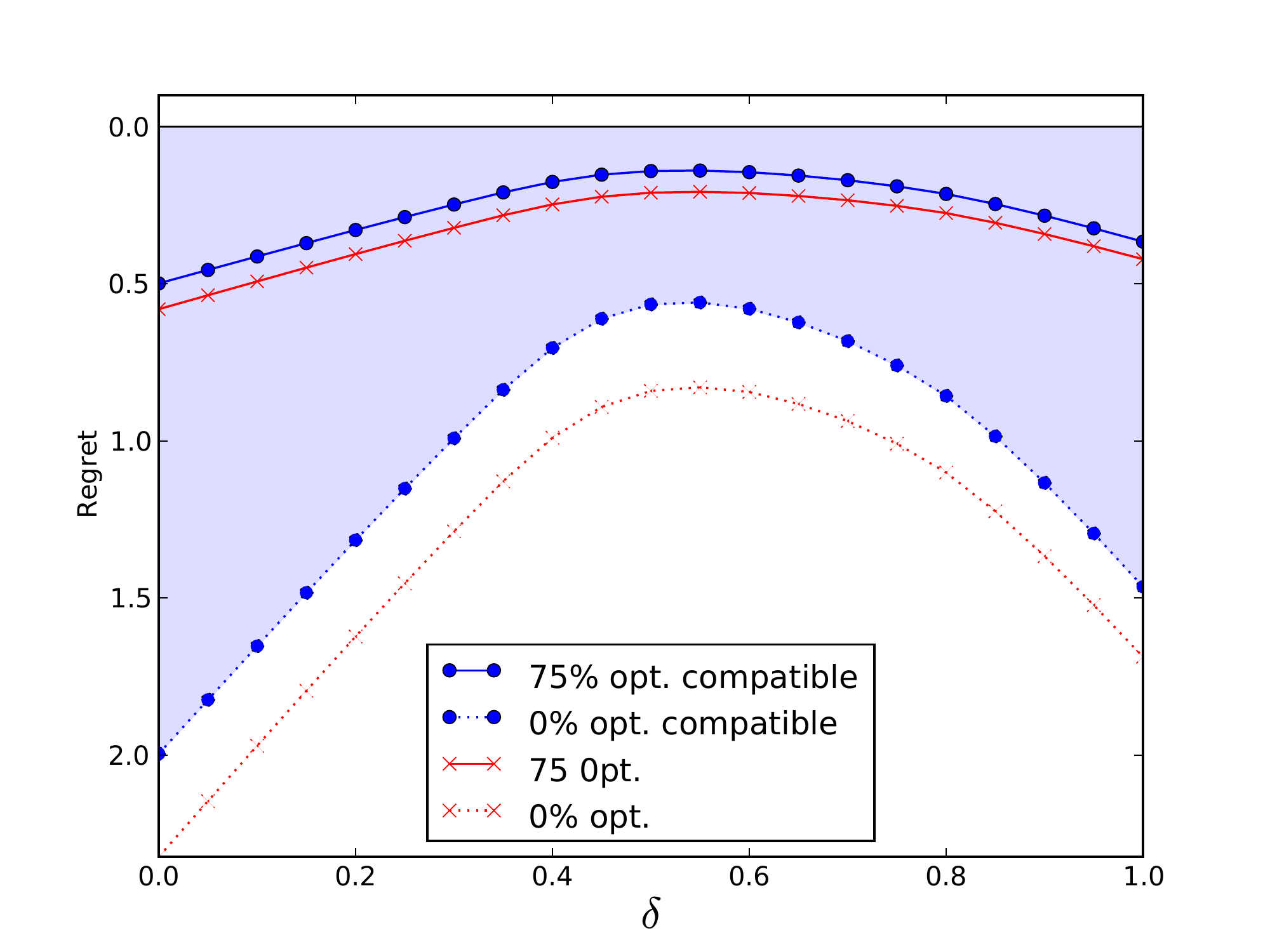}
\end{minipage}
\hspace{.05\linewidth}
\begin{minipage}{.28\linewidth}
\includegraphics[width=2.5in]{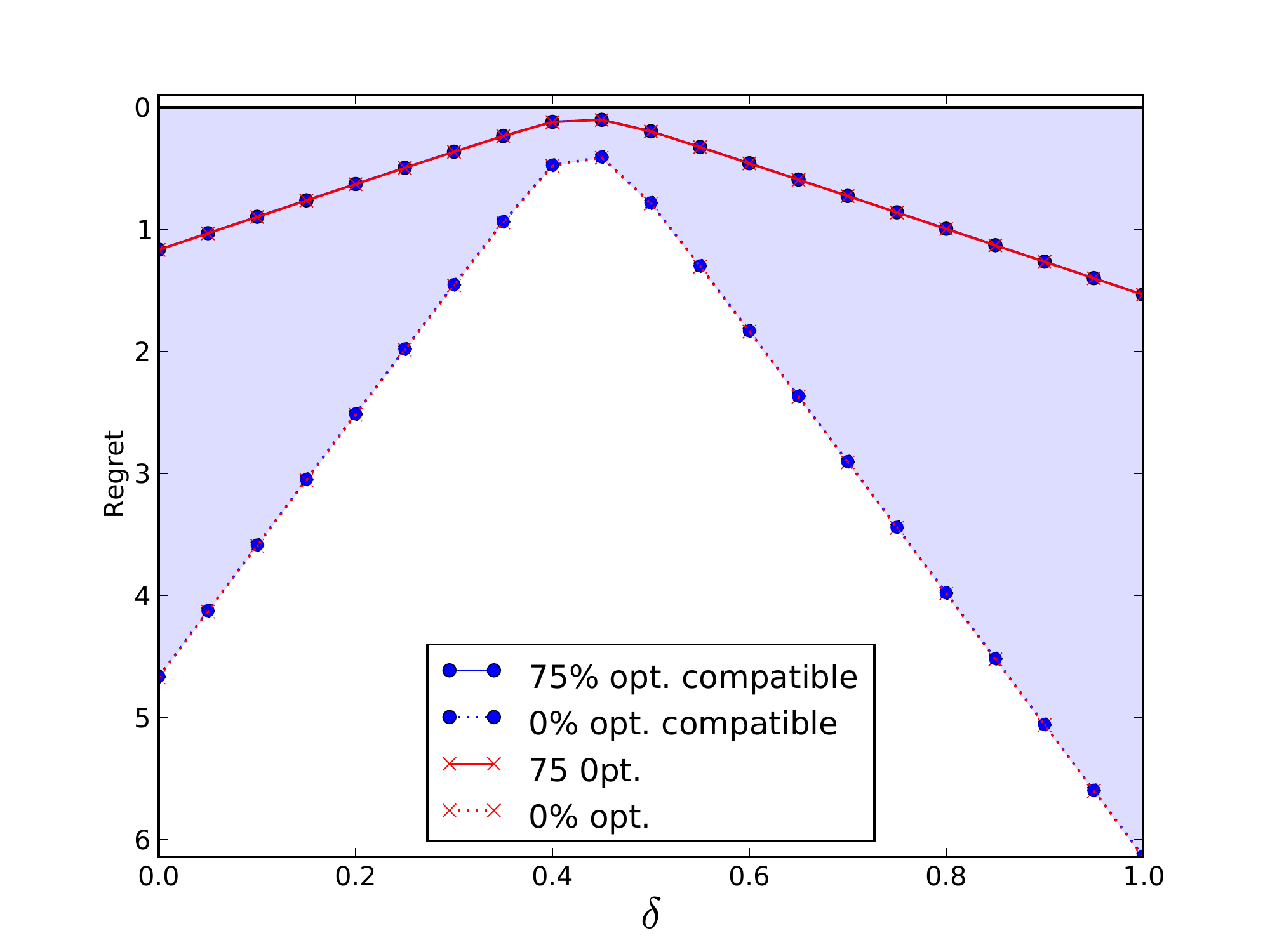}
\end{minipage}
\label{redVBluePlots}
\caption{Regret versus patient preference for the 75\% and 0\% consistent
  policies as well as $\pi_{c,0.5}$ and $\pi_{c,0.25}$.  The blue shaded
  region indicates the space of all policies that are compatible with the
  set-valued decision rule.  From left to right: parameter settings 1, 2, and
  3.}
\end{figure}

Figure \ref{greenVBluePlots} compares the regret of composite-outcome
based policies $\pi_{c, 0.5}$ and $\pi_{c, 0.25}$ with
set-valued-policy-derived 75\% and 0\% optimal compatible policies
across a range of patient preferences.  When there is a small fraction
of individuals with significant treatment effects running in opposite
directions (e.g., setting 1, when \texttt{Opst} is small) then the
composite outcome based policies perform well unless the preference is
grossly misspecified.  However, when there is a moderate to large
fraction of individuals with significant treatment effects running in
opposing directions (e.g., settings 2 and 3, when \texttt{Opst} is
moderate and large) then using a composite outcome based on only a
slightly misspecified patient preference can lead to regrets near the
0\% optimal compatible policy!  In contrast, the 75\% optimal
compatible policy remains relatively stable across all three settings
and all patient preferences even though no patient preference
information is required to estimate the set-valued decision rule.
Figure \ref{redVBluePlots} compares the regret of the 75\% and 0\%
optimal policies with the 75\% and 0\% optimal compatible policies.
When there are many individuals with unique best treatments (e.g.,
setting 1, when \texttt{Uniq} is large) then the set-valued decision
rule screens many suboptimal treatments and we see that the 75\%
optimal policy is dominated by the 0\% optimal compatible policy.  As
there are fewer individuals with unique optimal treatments the
difference between the 75\% optimal compatible and 75\% optimal
policies as well as the difference between the 0\% optimal compatible
and 0\% optimal policies converges to zero.  The reason for this is
clear, as there are a smaller number of unique best treatments fewer
suboptimal treatments are screened out.

%\begin{itemize}
%  \item Derive set-valued decision rule for a single stage 
%    decision problem with binary treatments.
%  \item Notes: practical significance levels $\Delta_{Y}$ and
%    $\Delta_{Z}$ can be elicited from clinicians or estimated 
%    from historical data if longitudinal data are available (e.g.
%    at what change in QIDS score did the clinician decide to change
%    treatment).  
%\end{itemize}

\section{Dynamic set-valued decision rules}\label{ss:twopoint}
In this section we extend set-valued decision rules to the case with
two decision points and two competing outcomes.  In order to make this
extension we will need to generalize the notation from the previous
section.  The data available estimate a pair of decision rules, one
for each time point, is denoted by $\mathcal{D} = \lbrace (H_{1,i},
A_{1,i}, H_{2,i}, A_{2,i}, Y_{i}, Z_{i})\rbrace_{i=1}^n$ and is comprised of
trajectories $(H_1, A_1, H_2, A_2, Y)$ drawn independently from a fixed but
unknown distribution.  The elements in each trajectory are as follows:
$H_t \in\mathbb{R}^{p_t}$, $t=1,2$ denotes the patient information available
to the decision maker \textit{before} the $t$th decision point\footnote{Note
that $H_2$ may contain some or all of the vector $(A_1,
H_1^{\T})^{\T}$.}; $A_t\in\lbrace -1,1\rbrace$, for $t=1,2$ denotes the
randomly assigned treatment at the $t$th stage; $Z,Y\in \mathbb{R}$
denotes competing outcomes observed sometime after the assignment of the
second treatment $A_2$. As in the previous section, we assume that both $Y$
and $Z$ are coded so that higher values are preferred.  Also available
are the quantities $\Delta_Y$ and $\Delta_Z$ denoting clinically relevant
quantities for $Y$ and $Z$ respectively.  
  
The goal is to construct a pair of decision rules $\pi = (\pi_1, \pi_2)$
where $\pi_t:\mathbb{R}^{p_t}$ $\rightarrow \lbrace \lbrace -1,1\rbrace,
\lbrace -1 \rbrace, \lbrace 1\rbrace \rbrace$ maps up-to-date patient
information to a subset of the possible decisions.  Like the ideal
static decision rule considered in the previous section, for 
a patient with second stage history $h_2$, the ideal
second stage decision rule, $\pi_{2\Delta}^\ideal$, should recommend a single
treatment if that treatment is expected to yield a clinically meaningful 
improvement in one or both of the outcomes without leading to 
significant loss in either.  Thus, by straightforward extension of the 
notation for the static-decision case we have, 
\begin{equation}\label{piOneStageIdeal}
\pi_{2\Delta}^{\ideal}(h_2) = \left\lbrace
\begin{array}{l}
\lbrace\sgn(r_{2Y}(h_2))\rbrace,\,\,\mathrm{if}\,\, 
|r_{2Y}(h_2)| \ge \Delta_{Y}\,\,
\mathrm{and}\,\,
\sgn(r_{2Y}(h_2))r_{2Z}(h_2) > -\Delta_Z, \\ 
\lbrace\sgn(r_{2Z}(h_2))\rbrace,\,\,\mathrm{if}\,\, |r_{2Z}(h_2)| 
\ge \Delta_Z\,\,
\mathrm{and}\,\,
\sgn(r_{2Z}(h_2))r_{2Y}(h_2) > -\Delta_Y, \\
\lbrace -1, 1\rbrace, \,\,\mathrm{otherwise,}\,\,
\end{array}
\right.
\end{equation}
where $r_{2Y}(h_2) \triangleq \mathbb{E}(Y|H_2 = h_2, A_2 = 1) - 
\mathbb{E}(Y|H_2 = h_2, A_2 = -1)$ and similarly,
$r_{2Z}(h_2) \triangleq \mathbb{E}(Z|H_2 = h_2, A_2 = 1) - \mathbb{E}(Z|H_2 =h _2, A_2 =-1)$.   

  We now define $\pi^\ideal_{1\Delta}$ given that a clinician
  always selects treatments from the set-valued decision rule
  $\pi^\ideal_{2\Delta}$ at the second stage. This problem is complicated by
  the fact that, unlike in the standard setting, there exists a set of
  histories $h_2$ at the second stage---those for which
  $\pi^\ideal_{2\Delta}(h_2) = \{-1,1\}$---where we do not know the
  treatment that would be chosen. To address this, we begin by
  assuming we know that some particular {\it non}-set-valued decision
  rule $\tau_2$ will be used at the second stage; we will then
  consider an appropriate set of possible $\tau_2$ in order to define
  $\pi^\ideal_{1\Delta}$.

Suppose that a clinician uses the non-set-valued decision rule $\tau_2:\mathbb{R}^{p_2}\rightarrow
\lbrace -1 ,1 \rbrace$ to assign treatments to patients at the second
decision point.  That is, if a patient presents with history $h_2$
then the clinician will assign treatment $\tau_2(h_2)$.  What is the
ideal decision rule at the first decision point knowing that the
clinician is using $\tau_2$ to assign treatments at the second
decision point?  For any function $\tau_2:\mathbb{R}^{p_2} \rightarrow
\lbrace -1, 1\rbrace$ define $Q_{2Y}(h_2, \tau_2) \triangleq
\mathbb{E}(Y|H_2 =h_2, A_2 = \tau_2(h_2))$.  Furthermore, define
$Q_{1Y}(h_1, a_1, \tau_2) \triangleq \mathbb{E}(Q_{2Y}(H_2,
\tau_2)|H_1 = h_1, A_1 = 1)$ so that $Q_{1Y}(h_1, a_1, \tau_2)$ is the
expected outcome for a patient with first stage history $H_1 = h_1$
treated at the first decision point with $A_1 = a_1$ and the decision
rule $\tau_2$ at the second decision point.  Replacing $Y$ with $Z$
above gives the definitions of $Q_{2Z}(h_2, \tau_2)$ and $Q_{1Z}(h_1,
a_1, \tau_2)$.  Thus, if it is known that a clinician will follow 
$\tau_{2}$ at the second decision point, then the ideal decision rule
at the first decision point is given by
\begin{equation}\label{piOneStageIdeal}
\pi_{1\Delta}^{\ideal}(h_1, \tau_2) = \left\lbrace
\begin{array}{l}
\lbrace\sgn(r_{1Y}(h_1, \tau_2))\rbrace,\,\,\mathrm{if}\,\, 
|r_{1Y}(h_1, \tau_2)| \ge \Delta_{Y}\,\,
\mathrm{and}\,\,
\sgn(r_{1Y}(h_1, \tau_2))r_{1Z}(h_1, \tau_2) > -\Delta_Z, \\ 
\lbrace\sgn(r_{1Z}(h_1, \tau_2))\rbrace,\,\,\mathrm{if}\,\, |r_{1Z}(h_1,
\tau_2)| 
\ge \Delta_Z\,\,
\mathrm{and}\,\,
\sgn(r_{1Z}(h_1, \tau_2))r_{1Y}(h_1,\tau_2) > -\Delta_Y, \\
\lbrace -1, 1\rbrace, \,\,\mathrm{otherwise,}\,\,
\end{array}
\right.
\end{equation}
where $r_{1Y}(h_1, \tau_2) \triangleq Q_{1Y}(h_1, 1, \tau_2) -
Q_{1Y}(h_1, -1, \tau_2)$, and similarly $r_{1Z}(h_1, \tau_2) \triangleq
Q_{1Z}(h_1, 1, \tau_2) - Q_{1Z}(h_1, -1, \tau_2)$.  Note that
$\pi_{1\Delta}^{\ideal}(h_2, \tau_2)$ assigns a single
treatment if that treatment is expected to yield a clinically
meaningful improvement on one or both the outcomes while not causing
clinically meaningful loss in either outcome whilst \textit{accounting
  for the clinician's behavior at the second decision point}, assuming
that behaviour is described by the non-set-valued rule $\tau_2$.  

  We now describe how to construct the ideal decision rule
  at the first decision point when the rule at the second decision
  point is set-valued. We say a \textit{non}-set-valued rule $\tau_2$ is
  \textit{compatible} with a set-valued decision rule $\pi_2$
  if and only if
\begin{equation}\label{eq:compatible}
  \tau_2(h_2) \in \pi_2(h_2)~~\forall h_2 \in \mathbb{R}^{p_2}.
\end{equation}
Let $\mathcal{C}(\pi_{2\Delta}^{\ideal})$ be the set of all rules that are
compatible with $\pi_{2\Delta}^{\ideal}$. We define 
$\pi_{1\Delta}^{\ideal}$ to be a
set-valued decision rule
\begin{equation}\label{eq:piTwoStageIdealSet}
\pi_{1\Delta}^{\ideal}(h_1) = \bigcup_{\tau \in
  \mathcal{C}(\pi_{2\Delta}^{\ideal})} \pi_{1\Delta}^\ideal(h_1,\tau).
\end{equation}
Our motivation for this definition is driven by a desire to maintain
as much choice as possible at stage 1, while making as few assumptions
about future behaviour as possible. We assume only that in the future
some $\tau_2$ in accordance with $\pi_2$ will be followed. Therefore at
stage 1 we only eliminate actions for which there exists \textit{no
 compatible future decision rule} that makes that action a desirable choice.

Note that if we do not assume a particular functional form for $\tau$,
the set $\mathcal{C}(\pi_{2\Delta}^{\ideal})$ will be very large, and
computing the union (\ref{eq:piTwoStageIdealSet}) will be
intractable. In practice, we will see that the modelling choices made
in order to estimate $Q_{2Y}$ and $Q_{2Z}$ suggest a reasonable subset
of $\mathcal{C}(\pi_{2\Delta}^{\ideal})$ over which we will take the union
(\ref{eq:piTwoStageIdealSet}) instead. We will provide a mathematical
programming formulation that allows us to use existing optimization
algorithms to efficiently compute the union over this smaller subset.

We now turn to the estimation of $\pi_{1\Delta}^\ideal$ and
$\pi_{2\Delta}^\ideal$ from data.  First, we note that to estimate the
ideal second decision rule we simply apply the results for the static
set-valued decision rule developed in the previous section.  That is,
we postulate linear models for second stage $Q$-functions, say, of the
form
\begin{eqnarray*}
Q_{2Y}(h_2, a_2) &=& h_{2,1}^{\T}\beta_{2Y} + a_2h_{2,2}^{\T}\psi_{2Y}, \\
Q_{2Z}(h_2, a_2) &=& h_{2,1}^{\T}\beta_{2Z} + a_2h_{2,2}^{\T}\psi_{2Z},
\end{eqnarray*}
which we estimate using least squares.  The estimated ideal second stage
set-valued decision rule $\hat{\pi}_{2\Delta}$ takes the form given in
(\ref{staticOLSOpt}).  In order to estimate the ideal decision rule at
the first decision point we must characterize how a clinician might
assign treatments at the second decision point.  We make the
assumption that clinicians' behavior, denoted $\tau_2$, is {\em
  compatible} with $\hat{\pi}_{2\Delta}$ as defined in
(\ref{eq:compatible}), and we assume that $\tau_2$ can be expressed as
a thresholded linear function of $h_2$. We call such decision rules
{\em feasible} for $\hat{\pi}_{2\Delta}$, and we define the set of
feasible decision rules at stage 2 by
\begin{equation*}\label{eq:feasible}
\mathcal{F}(\hat{\pi}_{2\Delta}) \triangleq 
\left\lbrace
\tau_{2}: \exists \rho\in\Re^{p_2}~\mathrm{s.t.}~\tau_2(h_2) = \sgn(h_{2,2}^{\T}\rho)~\mathrm{and}~\tau_2 \in \mathcal{C}(\hat{\pi}_{2\Delta})
\right\rbrace.
\end{equation*}

Thus, $\mathcal{F}(\hat{\pi}_{2\Delta})$ denotes the collection of
second stage non-set-valued decision rules that a clinician might
follow if they were presented with $\hat{\pi}_{2\Delta}$.  Note that
$\mathcal{F}(\hat{\pi}_{2\Delta})$ is non-empty since
$\sgn(h_{2,2}^{\T}(\frac{1}{2\Delta_Y}\hat{\psi}_{2Y} +
\frac{1}{2\Delta_Z}\hat{\psi}_{2Z}))$ belongs to
$\mathcal{F}(\hat{\pi}_{2\Delta})$.  For
an arbitrary $\tau_2 \in \mathcal{F}(\hat{\pi}_{2\Delta})$ define the
working models
\begin{eqnarray}
Q_{1Y}(h_1, a_1, \tau_2) &=& h_{1,1}^{\T}\beta_{1Y}(\tau_2) + 
a_1h_{1,2}^{\T}\psi_{1Y}(\tau_2),  \nonumber \\
Q_{1Z}(h_1, a_1, \tau_2) &=& h_{1,1}^{\T}\beta_{1Z}(\tau_2) + 
a_1h_{1,2}^{\T}\psi_{1Z}(\tau_2), \label{dynamicQ1WorkingModels}
\end{eqnarray}
where $\beta_{1Y}(\tau_2), \psi_{1Y}(\tau_2), \beta_{1Z}(\tau_2)$, and
$\psi_{1Z}(\tau_2)$ are coefficient vectors specific to $\tau_2$.  For
a fixed $\tau_2$ one can estimate these coefficients by regressing
$\hat{Q}_{2Y}(H_2, \tau_2) = H_{2,1}^{\T}\hat{\beta}_{2Y} +
\tau_2(H_2)H_{2,2}^{\T}\hat{\psi}_{2}$ and $\hat{Q}_{2Z}(H_2, \tau_2) =
H_{2,1}^{\T}\hat{\beta}_{2Z} + \tau_2(H_2)H_{2,2}^{\T}\hat{\psi}_{2Z}$ on
$H_1$ and $A_1$ using the working models in
(\ref{dynamicQ1WorkingModels}).  Let $\hat{Q}_{1Y}(h_1, a_1, \tau_2)$
and $\hat{Q}_{1Z}(h_1, a_1, \tau_2)$ denote these fitted models, and
let $\hat{r}_{1Y} \triangleq \hat{Q}_{1Y}(h_1, 1, \tau_2) -
\hat{Q}_{1Y}(h_1, -1, \tau_2)$, and $\hat{r}_{1Z} \triangleq
\hat{Q}_{1Z}(h_1, 1, \tau_2) - \hat{Q}_{1Z}(h_1, -1, \tau_2)$. We then
define
\begin{equation}\label{piOneStageIdeal}
\hat{\pi}_{1\Delta}^{\ideal}(h_1, \tau_2) = \left\lbrace
\begin{array}{l}
\lbrace\sgn(\hat{r}_{1Y}(h_1, \tau_2))\rbrace,\,\,\mathrm{if}\,\, 
|\hat{r}_{1Y}(h_1, \tau_2)| \ge \Delta_{Y}\,\,
\mathrm{and}\,\,
\sgn(\hat{r}_{1Y}(h_1, \tau_2))\hat{r}_{1Z}(h_1, \tau_2) > -\Delta_Z, \\ 
\lbrace\sgn(\hat{r}_{1Z}(h_1, \tau_2))\rbrace,\,\,\mathrm{if}\,\, |\hat{r}_{1Z}(h_1,
\tau_2)| 
\ge \Delta_Z\,\,
\mathrm{and}\,\,
\sgn(\hat{r}_{1Z}(h_1, \tau_2))\hat{r}_{1Y}(h_1,\tau_2) > -\Delta_Y, \\
\lbrace -1, 1\rbrace, \,\,\mathrm{otherwise,}\,\,
\end{array}
\right.
\end{equation}
and we define
\begin{equation}\label{eq:pi1hat}
\hat{\pi}_{1\Delta}(h_1) = \bigcup_{\tau_2 \in \mathcal{F}(\hat{\pi}_{2\Delta})}\hat{\pi}_{1\Delta}(h_1,\tau_2).
\end{equation}

\begin{comment}
\begin{equation}\label{piOneStageIdeal}
\hat{\pi}_{1\Delta}^{\inf}(h_1) = \left\lbrace
\begin{array}{ll}
\lbrace\sgn(h_1^{\T}\hat{\psi}_{1Y}(\tau_2)\rbrace,\,\,\mathrm{if}\,\, & 
\inf_{\delta_2\in\mathcal{F}(\hat{\pi}_{2\Delta})}|2h_1^{\T}\hat{\psi}_{1Y}(\delta_2)| 
\ge \Delta_{Y}\,\,
\mathrm{and}\,\, \\ & 
\inf_{\delta_2\in\mathcal{F}(\hat{\pi}_{2\Delta})}\sgn(h_1^{\T}\hat{\psi}_{1Y}(\tau_2))2h_1^{\T}\hat{\psi}_{1Z}(\delta_2)
 > -\Delta_Z, \\
\lbrace\sgn(h_1^{\T}\hat{\psi}_{1Z}(\tau_2)\rbrace,\,\,\mathrm{if}\,\, & 
\inf_{\delta_2\in\mathcal{F}(\hat{\pi}_{2\Delta})}|2h_1^{\T}\hat{\psi}_{1Z}(\delta_2)| 
\ge \Delta_{Z}\,\,
\mathrm{and}\,\, \\ & 
\inf_{\delta_2\in\mathcal{F}(\hat{\pi}_{2\Delta})}\sgn(h_1^{\T}\hat{\psi}_{1Z}(\tau_2))2h_1^{\T}\hat{\psi}_{1Y}(\delta_2)
 > -\Delta_Y, \\ 
\lbrace -1, 1\rbrace & \mathrm{otherwise},
%\lbrace\sgn(r_{1Z}(h_1, \tau_2))\rbrace,\,\,\mathrm{if}\,\, r_{1Z}(h_1,
%\tau_2) 
%\ge \Delta_Z\,\,
%\mathrm{and}\,\,
%\sgn(r_{1Z}(h_1, \tau_2))r_{1Y}(h_1,\tau_2) > -\Delta_Z, \\
%\lbrace -1, 1\rbrace, \,\,\mathrm{otherwise,}\,\,
\end{array}
\right.
\end{equation}
where $\tau_2 \in \mathcal{F}(\hat{\pi}_{2\Delta})$ is arbitrary.  
\end{comment}

Thus, $\hat{\pi}_{1\Delta}$ is a set-valued decision rule that assigns
a single treatment if only that treatment leads to an (estimated)
expected clinically meaningful improvement on one or both outcomes and
does not lead to a clinically meaningful loss in either outcome across
all the treatment rules in $\mathcal{F}(\hat{\pi}_{2\Delta})$ that a
clinician might consider at the second stage.  Alternatives to this
definition of $\hat{\pi}_{1\Delta}$ are discussed in Section 6.
\subsection{Computation}\label{ss:computation}
Computing $\hat{\pi}_{1\Delta}(h_1)$ requires solving what appears to
be a difficult enumeration problem.  
%In particular, if
%$\delta_2(h_2) = \sgn(h_2^{\T}\eta)$ then
%$\hat{\psi}_{2Y}(\delta_2)$ is a non-smooth function of $\eta$.
%Furthermore, the set $\mathcal{F}(\hat{\pi}_{2\Delta})$ will typically
%be infinite (one exception occurs when $\hat{\psi}_{2Y} =
%\hat{\psi}_{2Z}$, but in this degenerate case set-valued regimes are
%not needed).  
%% -- I don't get the above; \mathcal{F} is a set of set-valued policies
%% right? Aren't there 3^n of these?
Nevertheless, exact computation of
$\hat{\pi}_{1\Delta}(h_1)$ is possible and (\ref{eq:pi1hat}) can be solved
quickly when $p_2$ is small.  

First, note that if $\tau_2$ and $\tau'_2$ are decision rules at the
second stage that agree on the observed data $\mathcal{D}$, that is,
$\tau_2(h_{2i}) = \tau'_2(h_{2i})$ for all values $h_{2i}$ in
$\mathcal{D}$, then $\hat{\psi}_{1Y}(\tau_2) =
\hat{\psi}_{1Y}(\tau'_2)$ and $\hat{\psi}_{1Z}(\tau_2) =
\hat{\psi}_{1Z}(\tau'_2)$. It follows that
$\hat{\pi}_{1\Delta}(h_1,\tau_2) =
\hat{\pi}_{1\Delta}(h_1,\tau'_2)~~\forall h_1\in H_1$. Thus, if we
consider a finite subset $\tilde{\mathcal{F}}(\hat{\pi}_{2\Delta})$ of
$\mathcal{F}(\hat{\pi}_{2\Delta})$ that contains one decision rule for
each possible ``labeling'' of the stage 2 histories contained in
$\mathcal{D}$, then we have 
\begin{equation}
\hat{\pi}_{1\Delta}(h_1) = \bigcup_{\tau_2\in
  \mathcal{F}(\hat{\pi}_{2\Delta})}\hat{\pi}_{1\Delta}(h_1,\tau_2) =
\bigcup_{\tau_2\in
  \tilde{\mathcal{F}}(\hat{\pi}_{2\Delta})}\hat{\pi}_{1\Delta}(h_1,\tau_2).
\end{equation}
We use the term ``labeling'' by analogy with classification:
histories at stage 2 are given a binary ``label'' by $\tau_2$ which is
either $1$ or $-1$. Rather than taking a union over the potentially
uncountable $\mathcal{F}(\hat{\pi}_{2\Delta})$, we will enumerate the
finite set of all feasible {\em labelings} of the observed data,
place a corresponding $\tau_2$ for each one into the set
$\tilde{\mathcal{F}}(\hat{\pi}_{2\Delta})$, and take the union over
$\tilde{\mathcal{F}}(\hat{\pi}_{2\Delta})$ instead.

We say that a labeling $\ell_i, i=1,\ldots,n$ is
\textit{compatible} with a set-valued decision rule
$\hat{\pi}_{2\Delta}$ if $\ell_i \in \hat{\pi}_{2\Delta}(h_{2,i}), i =
1,\ldots,n$ and \textit{feasible} if it can be induced by a feasible
decision rule $\tau_2 \in
\mathcal{F}(\hat{\pi}_{2\Delta})$. Equivalently, the labeling is
feasible if it is both compatible with $\hat{\pi}_{2\Delta}$ and if
the two sets $\{h_{2i} | \ell_i = 1\}$ and $\{h_{2i} | \ell_i = -1\}$
are {\em linearly separable} in $\Re^{p_2}$. Our algorithm for
computing $\tilde{\mathcal{F}}(\hat{\pi}_{2\Delta})$ works by
specifying a mixed integer linear program with indicator
constraints\footnote{In the field of mathematical programming, the
  term ``indicator constraint'' is used for a constraint that is only
  enforced when a variable takes on a particular value, e.g.\ when an
  indicator variable is 1. (A better term might be ``conditional
  constraint.'')} whose feasible solutions correspond to the linearly
separable labelings of $\mathcal{D}$ that are compatible with
$\hat{\pi}_{2\Delta}$.

First, we note that determining whether or not a given labeling
$\ell_i, i \in 1,\ldots,n$ is compatible with $\hat{\pi}_{2\Delta}$ is
equivalent to checking the following set of constraints:
\begin{equation}\label{eq:compatiblelinear}
\exists \psi_2~\mathrm{s.t.}~\ell_i h_{2,i}^\T\psi_2 \ge 1~~\forall i \in 
1,\ldots, n
\end{equation}
Given a particular labeling, the existence of a $\psi_2$ that
satisfies (\ref{eq:compatiblelinear}) can be proven or disproven in
polynomial time using a linear program solver (see, for example, Megiddo 1987
and references therein). The
existence of such a $\psi_2$ implies a compatible $\tau$ given by
$\tau(h_2) = \sgn (h_2^\T \psi_2)$ that produces labeling
$\ell_{1}, \ldots, \ell_{n}$ when applied to
the stage 2 data.

To find all possible feasible labelings,
we formulate the following mixed integer linear program with indicator constraints
\begin{align*}
\min_{\ell_1,\ell_2,...,\ell_n,\psi_2} & f(\ell_1,\ell_2,...,\ell_n,\psi_2) \\
\mathrm{s.t.}~\forall i \in 1,\ldots, n,~& \ell_i \in \{-1,1\}\\
&
\begin{array}{ll} 
h_{2,2,i}^\T \psi_2 \ge 1 & \mbox{if $\ell_i = 1$}\\
h_{2,2,i}^\T \psi_2 \le -1 & \mbox{if $\ell_i = -1$}
\end{array}
\\
& \psi_2 \in \Re^{p_2}\\
\end{align*}
and find all feasible solutions.  Note that exactly one of the
constraints involving $h_{2i}$ is enforced for a particular value of
$\ell_i$. We present the feasibility problem as a minimization because
it is the natural form for modern optimization software packages like
CPLEX, which are capable of handling both the integer constraints on
$\ell_i$ and the indicator constraints on $\psi_2$. Note that if we
simply want to recover the feasible $\ell_i$ then the choice of $f$
does not matter, and we may choose $f = 0$ for simplicity and
efficiency in practice.  CPLEX is capable of enumerating \textit{all} 
feasible solutions efficiently (the examples considered in this 
paper take less than one minute to run on a laptop
with 8GB DDR3 RAM and a 2.67GHz dual core processor).  
If we wanted to also recover the maximum
margin separators for the feasible labelings, for example, we could
use the quadratic objective $f = \|\psi_2 \|^2$.

Let $\tilde{\mathcal{F}}(\hat{\pi}_{2\Delta})$ denote the collection
of feasible decision rules defined by $\sgn (h_{2,2}^{\T}\psi_2)$ for each
feasible $\psi_2$. Then for any $h_1 \in \mathbb{R}^{p_1}$
\begin{equation}
\hat{\pi}_{1\Delta}(h_1) = \bigcup_{\tau_2 \in
  \tilde{\mathcal{F}}(\hat{\pi}_{2\Delta})}\hat{\pi}_{1\Delta}(h_1,\tau_2).
\end{equation}
Note that $\tilde{\mathcal{F}}(\hat{\pi}_{2\Delta})$ does not depend
on the vector $h_1$ and hence only needs to be computed once for a
given dataset.

%\begin{itemize}
%  \item Derive the set-valued decision rule for the two stage 
%    binary treatment case.  In this development we will use the 
%    $\sup/\inf$ formulation.  
%  \item Discuss alternatives to the $\sup/\inf$ formulation, e.g.\ 
%    recommend a set if more than 5\% of the feasible policies at the
%    second stage would lead to recommending a set.
%  \item Discuss computational issues.  
%\end{itemize}

\section{Examples}\label{Examples}
\subsection{Nefazodone study}
In this section we illustrate the estimation of a static set-valued
decision rule for a single decision point with two competing outcomes.
The data are from the initial (12 week) phase of a multicenter
longitudinal study comparing three treatment combinations for chronic
depression \citep{keller2000comparison}.  A total of 681 subjects were randomly assigned with equal
probability to nefazodone only (Drug), cognitive behavioral therapy
only (CBT), or a combination of the two (Drug+CBT).  CBT requires
up to twice-weekly on-site visits to the clinic and thus, relative to Drug, CBT
represents a substantial time and monetary burden on patients.  Hence, an
important question is which patients are likely to benefit from CBT
beyond Drug only on one or more outcomes.  To focus on this question
and simplify our exposition, we consider only the treatments Drug and
Drug+CBT.

The primary outcome for the study was depression as measured by the
24-item Hamilton Rating Scale for Depression (HRSD) where lower scores
signify a healthier state.  However, nefazodone is associated with
fatigue and lack of physical coordination and thus physical
functioning represents an important competing outcome to depression.
Physical functioning is quantified in this study using the physical
functioning factor score in the Medical Outcomes Study 36-Item Short
Form Health Survey (PF).   PF was measured at baseline and at
12 weeks and we let $Z$ denote the 12 week measurement.  
HRSD was measured at baseline and weeks 1, 2, 3, 4, 6, 8, 10, and 12.  
Let $Y_j$ denote a patients HRSD at week $j$.
To reduce variability and to 
capture patient improvement over the duration of the  study, we 
define the outcome  $Y$ to be the least squares slope
of $Y_{0}, Y_{1}, \ldots, Y_{12}$ on the observation times
$j=0, 1, \ldots, 12$.  For patients missing HRSD measurements $Y$ was computed
using the least squares slope of the observed measurements on the 
observed measurement times.  

Of the $681$ patients enrolled in the study, $226$ were assigned
to Drug while $227$ were assigned to Drug+CBT.  
PF was not measured on all patients so
we use a subset of $164$ patients assigned to
Drug and $172$ patients with Drug+CBT with complete PF measurements.  There 
was no missingness in baseline covariates.  
% Because the HRSD was
% administered at multiple time points during the 12 week period, we
% chose to use rate of change (slope) of the subject's scores as the
% response variable instead of the final HRSD score.  We also
% transformed all of the scores so that higher scores represented more
% favorable outcomes.  The physical functioning outcome represents the
% subject's limitations with physical related activities due to health
% related problems.  It comes from the physical functioning factor score
% found in the Medical Outcomes Study 26-Item Short Form Health Survey
% (MOS-36) which was administered at the beginning and end of the 12
% week study.  It ranges from 0-100 with higher scores indicating less
% limitations due to health problems.  For both outcomes, subjects with
% excessive missing values were dropped from this analysis.  The total
% number of subjects we used was 336.
In order to estimate the ideal decision rule derived in Section 3, we
modeled the conditional expectations $ \mathbb{E}(Y|H=h, A=a)$ and $
\mathbb{E}(Z|H=h, A=a)$ using the working models of the form given in
(\ref{workingQModelsSetValuedI}) and
(\ref{workingQModelsSetValuedII}).  There were a large number of
covariates collected at baseline and we constructed our models based
on clinician input and exploratory data analysis.  Typical regression
diagnostics for linear regression ~\citep[e.g.,][]{cook99appliedregression}
suggest that the models fit the data reasonably well.  The
covariates included in the model for depression (outcome $Y$) were the
subject's baseline Hamilton depression score (\texttt{hamd}); role
functioning physical factor score (a measure of the ability to perform
physical related roles) (\texttt{rolfun}); the assigned treatment
(Drug was coded as -1 and Drug+CBT was coded as 1)(\texttt{trt}); and
the interactions \texttt{trt}*\texttt{hamd} and
\texttt{trt}*\texttt{rolfun}.  The covariates included in the physical
functioning model were the subject's baseline physical functioning
factor score (\texttt{phyfun}); patient gender (\texttt{gender});
sleep score (a measure of the subject's quality of sleep)
(\texttt{slpsc}); overall general health perception score
(\texttt{genhel}); role functioning score (\texttt{rolfun}); age of
onset of depression (\texttt{mdage}); presence of dysthemia
(\texttt{dyst}); the assigned treatment (\texttt{trt}), and the
interactions \texttt{trt}*\texttt{slpsc} and
\texttt{trt}*\texttt{phyfun}.  Tables 1 and
2 display the fitted parameters for each model.
%Note that the \texttt{hamd}*\texttt{trt} 
%interaction was not found
%significant in the depression model but we included it because we
%believed it should be in the model.

\begin{table}[H]
\begin{center}
\begin{tabular}{rrrrr}
  \hline
 & Estimate & Std. Error & t value & Pr($>$$|$t$|$) \\ 
  \hline
(Intercept) & 0.2867 & 0.2158 & 1.33 & 0.1848 \\ 
  \texttt{hamd} & 0.0325 & 0.0074 & 4.41 & 0.0000 \\ 
  \texttt{rolfun} & -0.0009 & 0.0009 & -1.01 & 0.3135 \\ 
  \texttt{trt} & 0.1133 & 0.2158 & 0.53 & 0.5998 \\ 
  \texttt{rolfun}*\texttt{trt} & 0.0018 & 0.0009 & 2.01 & 0.0452 \\ 
  \texttt{hamd}*\texttt{trt} & 0.0011 & 0.0074 & 0.14 & 0.8858 \\ 
   \hline
\end{tabular}
\label{depressionTable}
\caption{Summary of the fitted coefficients for depression model.}
\end{center}
\end{table}

\begin{table}[H]
\begin{center}
\begin{tabular}{rrrrr}
  \hline
 & Estimate & Std. Error & t value & Pr($>$$|$t$|$) \\ 
  \hline
(Intercept) & 35.6607 & 4.8319 & 7.38 & 0.0000 \\ 
  \texttt{gender} & -3.4435 & 1.5851 & -2.17 & 0.0305 \\ 
  \texttt{slpsc} & -0.0979 & 0.3481 & -0.28 & 0.7786 \\ 
  \texttt{phyfun} & 0.6198 & 0.0442 & 14.01 & 0.0000 \\ 
  \texttt{genhel} & 0.1384 & 0.0413 & 3.35 & 0.0009 \\ 
  \texttt{rolfun} & -0.0436 & 0.0206 & -2.12 & 0.0349 \\ 
  \texttt{mdage} & -0.1236 & 0.0583 & -2.12 & 0.0347 \\ 
  \texttt{dyst} & -3.8610 & 1.4859 & -2.60 & 0.0098 \\ 
  \texttt{trt} & 11.3895 & 3.8200 & 2.98 & 0.0031 \\ 
  \texttt{slpsc}*\texttt{trt} & -0.8737 & 0.3443 & -2.54 & 0.0116 \\ 
  \texttt{phyfun}*\texttt{trt} & -0.0714 & 0.0365 & -1.96 & 0.0511 \\ 
   \hline
\end{tabular}
\label{physfunTable}
\caption{Summary of the fitted coefficients for physical functioning model.}
\end{center}
\end{table}

\begin{figure}[H] %  figure placement: here, top, bottom, or page
   \centering
   \includegraphics[width=5in]{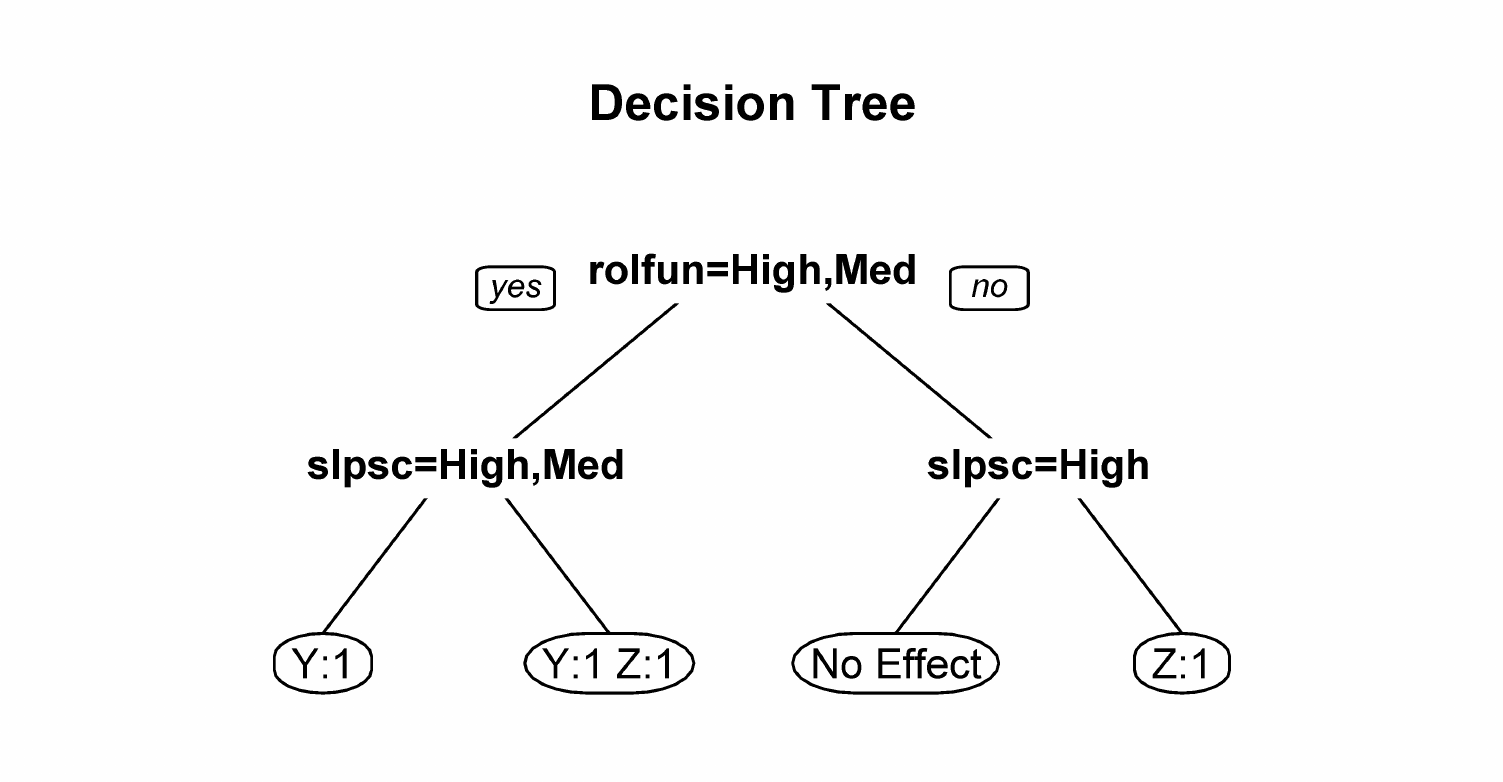} 
   \caption{A decision tree approximating the estimated ideal decision
     rule for the nefazodone study. The leaves of the tree indicate
     the optimal treatment (1 for combination or -1 for drug) along
     with which outcome (Y for depression or Z for physical
     functioning) the subject would see a clinically significant
     change in. }
   \label{bradTree}
\end{figure}

In what follows we use $\Delta_Y =.25$ as a clinically meaningful
difference for depression which translates into a change of 4 units in
a subject's depression score over 12 weeks.  HRSD is typically
categorized into one of five severity categories, four units roughly
corresponds to moving one of these categories.   
We use $\Delta_Z = 5$
which corresponds to a 5\% change on the scale PF is measured.  It is
important to note that these thresholds have been chosen here
primarily for illustrative purposes.  Using the foregoing values of
$\Delta_Y$ and $\Delta_Z$, we estimated the ideal decisions using the
methods described in Section 3.  To get a sense of the estimated
set-valued decision rule we approximated the estimated decision rule with a
decision tree.  Figure \ref{bradTree} displays the estimated ideal
decision rule as approximated using a decision tree.  Note,
`high',`medium' and `low' values were used instead of actual scores
for plot clarity.  Drug+CBT was always included in the set of
recommended treatments but for patients with low role functioning
(\texttt{rolfun}) and low sleep scores (\texttt{slpsc}) there a
negligible treatment effect and Drug only may not be significantly
different than Drug+CBT.  Another useful display of the set valued
decision rule is to plot the estimated contrasts $2h_{1,2}^{T}\hat{\psi}_Y$
and $2h_{1,2}^{T}\hat{\psi}_Z$ against each other and to label the regions
where the decision rule changes.  This plot is shown in Figure
\ref{fig:rtopiDEPRESSION}.

%\begin{figure}[H] %  figure placement: here, top, bottom, or page
%   \centering
%   \includegraphics[width=4in]{brad_YZGrid.pdf} 
%   \caption{Plot of fitted ideal decisions for each observation.  Note the lar%ge majority of subjects should be assigned treatment 1 (combination) with only% a few that should be assigned treatment -1 (drug).}
%   \label{bradGrid}
%\end{figure}

The primary analysis of the depression study found that the Drug+CBT
was the most effective ~\citep{keller2000comparison} and the results in
Figures \ref{bradTree} and \ref{fig:rtopiDEPRESSION} are consistent
with this analysis.
Almost always, the combination treatment was recommend when
considering depression and physical functioning as the competing
outcomes.  As seen in Figure \ref{bradTree}, only
\ref{fig:rtopiDEPRESSION} subjects were assigned the drug treatment as
their ideal treatment with the remaining subjects being assigned
either the combination treatment or no treatment.  Because of this,
the decision tree in Figure \ref{bradTree} only assigns future
subjects to the combination treatment or no treatment.  Additionally,
it provides information about which outcome the subject would likely
see clinically significant results in if they were to follow these
decisions.  Note that the \texttt{slpsc}*\texttt{trt} and 
\texttt{rolfun}*\texttt{trt}
interactions were both found significant in the models and it is at
these variables where the splits in the tree were made.

\newcommand{\Dy}{0.16666}
\newcommand{\Dz}{2.5}
\begin{figure}
\begin{center}
\begin{tikzpicture}[y=0.1cm,x=2.3cm,scale=6]

%Central region of indifference
%\pattern[pattern=dots,pattern color=lightgray] (-\Dy,-\Dz) rectangle (\Dy,\Dz);
%Lower-right
%\pattern[pattern=dots,pattern color=lightgray] (\Dy,-\Dz) rectangle (1,-1);
%Upper-left
%\pattern[pattern=dots,pattern color=lightgray] (-\Dy,\Dz) rectangle (-1,1);

\pgfmathparse{-2.5*\Dy}
\let\LeftEnd\pgfmathresult
\pgfmathparse{4*\Dy}
\let\RightEnd\pgfmathresult
\pgfmathparse{-2.5*\Dz}
\let\BottomEnd\pgfmathresult
\pgfmathparse{4*\Dz}
\let\TopEnd\pgfmathresult

%Prefer treatment 1
\pattern[pattern=vertical lines,pattern color=lightgray] (-\Dy,\Dz) rectangle (\Dy,\TopEnd);
\pattern[pattern=vertical lines,pattern color=lightgray] (\Dy,\Dz) rectangle (\RightEnd,\TopEnd);
\pattern[pattern=vertical lines,pattern color=lightgray] (\Dy,-\Dz) rectangle (\RightEnd,\Dz);

%Prefer treatment -1
\pattern[pattern=horizontal lines,pattern color=lightgray] (-\Dy,\Dz) rectangle (\LeftEnd,-\Dz);
\pattern[pattern=horizontal lines,pattern color=lightgray] (-\Dy,-\Dz) rectangle (\LeftEnd,\BottomEnd);
\pattern[pattern=horizontal lines,pattern color=lightgray] (\Dy,-\Dz) rectangle (-\Dy,\BottomEnd);

%Thick lines
\newcommand{\lnth}{2pt}
\draw[line width=\lnth] (-\Dy,-\Dz) rectangle (\Dy,\Dz);
\draw[line width=\lnth] (-\Dy,\Dz) -- (-\Dy,\TopEnd);
\draw[line width=\lnth] (-\Dy,\Dz) -- (\LeftEnd,\Dz);
\draw[line width=\lnth] (\Dy,-\Dz) -- (\Dy,\BottomEnd);
\draw[line width=\lnth] (\Dy,-\Dz) -- (\RightEnd,-\Dz);

%Y direction (horizontal)
\draw (\LeftEnd,0) -- (\RightEnd,0);

%Delta Z bounds
\draw[dotted] (\LeftEnd,\Dz) -- (\RightEnd,\Dz);
\draw[dotted] (\LeftEnd,-\Dz) -- (\RightEnd,-\Dz);

%Z direction (vertical)
\draw (0,\BottomEnd) -- (0,\TopEnd);

%Delta Y bounds
\draw[dotted] (\Dy,\BottomEnd) -- (\Dy,\TopEnd);
\draw[dotted] (-\Dy,\BottomEnd) -- (-\Dy,\TopEnd);

%Treatment labels
\node at (-\Dy,\Dz) [fill=white,anchor=south east,outer sep=10pt] {Output: $\{-1,1\}$};
\node at (0,0) [fill=white] {Output: $\{-1,1\}$};
\node at (\Dy,-\Dz) [fill=white,anchor=north west,outer sep=20pt] {Output: $\{-1,1\}$};

\node at (-\Dy,-\Dz) [fill=white,anchor=north east,outer sep=10pt] {Output: $\{-1\}$};
\node at (\RightEnd,\TopEnd) [fill=white,anchor=north east,outer sep=20pt] {Output: $\{1\}$};%

%Axis labels
\node at (0,\TopEnd) [anchor=south] {$r_Z(h)$};
\node at (\RightEnd,0) [anchor=west] {$r_Y(h)$};

%Delta Y ticks
\node at (\Dy,\TopEnd) [anchor=south] {$\Delta_Y$};
\node at (-\Dy,\TopEnd) [anchor=south] {$-\Delta_Y$};

%Delta Z ticks
\node at (\RightEnd,\Dz) [anchor=west] {$\Delta_Z$};
\node at (\RightEnd,-\Dz) [anchor=west] {$-\Delta_Z$};

\draw plot[only marks,mark=*,mark options={scale=0.05}] file {values_for_1stage_plot.txt};

\end{tikzpicture}

\end{center}
\caption{\label{fig:rtopiDEPRESSION}Diagram showing how the output of $\pi^\ideal(h)$ depends on
  $\Delta_Y$ and $\Delta_Z$, and on the location of the point $(r_Y(h),r_Z(h))$.}
\end{figure}
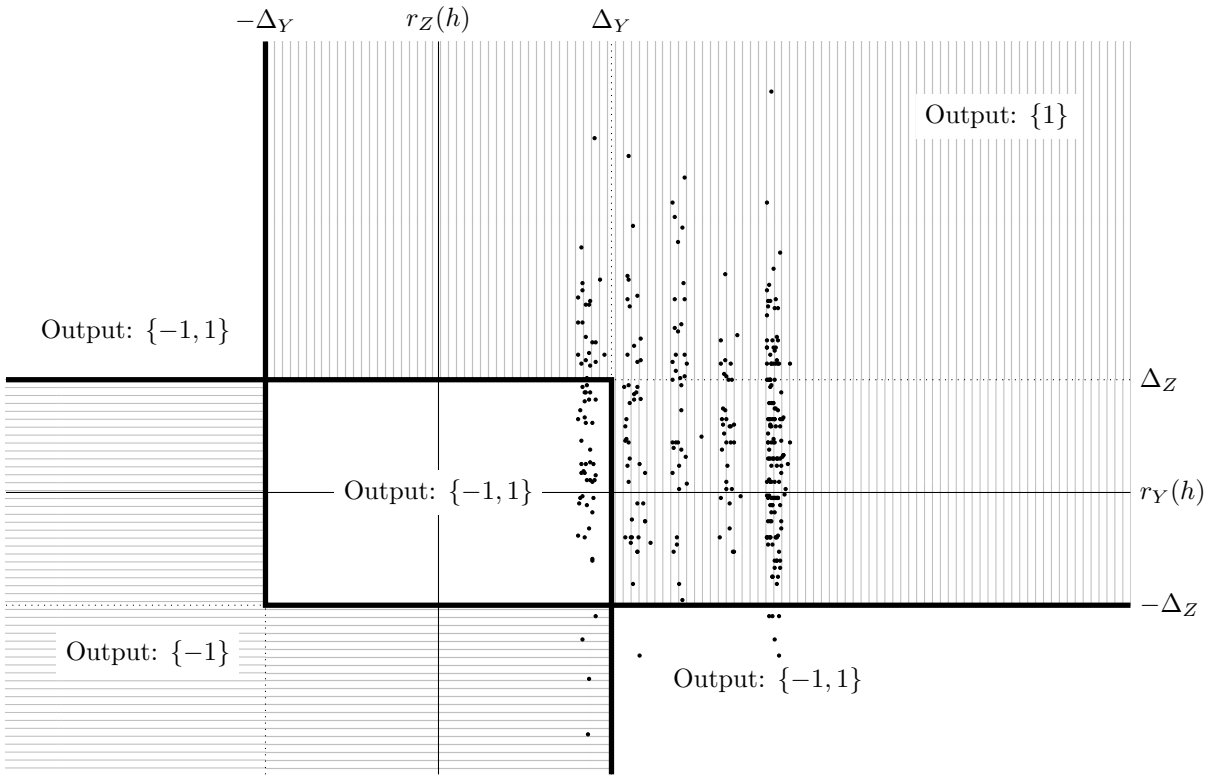

\subsection{CATIE}\label{ss:CATIE}
We now consider the application of our method to data from the
Clinical Antipsychotic Trials of Intervention Effectiveness (CATIE)
Schizophrenia study. The CATIE study was designed to compare sequences
of antipsychotic drug treatments for the care of schizophrenia
patients. The full study design is quite complex
~\citep{stroup03design}; we will make several simplifications in order
to more clearly illustrate the potential of the method presented
here. CATIE was an 18-month sequential randomized trial that was
divided into two main phases of treatment. Upon entry into the study,
most patients began ``Phase 1,'' in which they were randomized to one
of five possible treatments with equal probability: olanzapine,
risperidone, quetiapine, ziprasidone, or perphenazine. As they
progressed through the study, patients were given the opportunity at
each monthly visit to discontinue their Phase 1 treatment and begin
``Phase 2'' on a new treatment. The set of possible Phase 2 treatments
depended on the reason for discontinuing Phase 1 treatment. If the
Phase 1 treatment was deemed to produce unacceptable side-effects,
they entered the \textit{tolerability arm} and their Phase 2 treatment
was chosen uniformly randomly from the set \{olanzapine, risperidone,
quetiapine, ziprasidone\}.  If the Phase 1 treatment was deemed to be
ineffective at reducing symptoms, they entered the \textit{efficacy
  arm} and their Phase 2 treatment was chosen randomly as follows:
clozapine with probability $1/2$, or uniformly randomly from the set
\{olanzapine, risperidone, quetiapine\} with probability $1/2$.

Although CATIE was designed to compare several treatments within each
arm, there are natural groupings at each stage that allow us to
collapse the data in a meaningful way so that we consider only binary
treatments and we can therefore directly apply our method as
described. In the Phase 2 Tolerability arm, it is natural to compare
olanzapine against the other three drugs since it is known a priori to
be efficacious, but is also known to cause significant weight gain as
a side-effect. In the Phase 2 Efficacy arm, it is natural to compare
clozapine against the rest of the potential treatments, both because
the randomization probabilities called for having 50\% of patients in
that arm on clozapine, and because clozapine is substantively
different from the other three drugs: it is known to be highly
effective at controlling symptoms, but it is also known to have
significant side-effects and its safe administration requires very
close patient monitoring. In Phase 1, it is natural to compare
perphenazine, the only typical antipsychotic, against the other four
drugs which are atypical antipsychotics. (This comparison of
typical-versus-atypical was in fact an important goal of the CATIE
study.)

For our first outcome, $Y$, we use the Positive and Negative Syndrome
Scale (PANSS) which is a numerical representation of the level of
psychotic symptoms experienced by a patient ~\citep{kay87positive}. A
higher value of PANSS reflects the presence of more severe
symptoms. PANSS is a well-established measure that we have used in
previous work on the CATIE study ~\citep{shortreed11informing}. Since
having larger PANSS is worse, for our first outcome $Y$ we use 100
minus the percentile of a patient's PANSS at the end of their time in
the study. We use the distribution of PANSS at the beginning of the
study as the reference distribution for the percentile.

For our second outcome, we use Body Mass Index (BMI), a measure of
obesity. Weight gain is an important and problematic side-effect of
many antipsychotic drugs \citep{allison99antipsychotic}.  Since in
this population having a larger BMI is worse, for our second outcome
$Z$ we use 100 minus the percentile of a patient's BMI at the end of
their time in the study. Again, we use the distribution of BMI at the
beginning of the study as the reference distribution for the
percentile.

In all of our models, we include two baseline covariates. The first,
\texttt{td}, is a dummy variable indicating if a patient has ``tardive
dyskinesia,'' which is a motor side-effect that can be caused by
previous treatment. The second, \texttt{exacer}, is a dummy variable
indicating that a patient has been recently hospitalized, thus
indicating an exacerbation of his or her condition. These do not
interact with treatment.

For our covariates $h_2$ that interact with treatment, we choose the
patients most recently recorded PANSS score percentile in our model
for PANSS, and the most recently recorded BMI percentile in our model
for BMI. These percentiles were shifted by $-50$ so that a patient
with at the median has $h_2 = 0$. This was done so that in each model,
the coefficient for the main effect of treatment can be directly
interpreted as the treatment effect for a patient with median PANSS
(resp.\ BMI). Treatments were coded ${1,-1}$. For both outcomes we chose 5 percentile points as our
indifference range, so $\Delta_Y = \Delta_Z = 5$.

\subsubsection{Phase 2 Tolerability}

\begin{table}[H]
\begin{center}
\begin{tabular}{rrrrr}
  \hline
 & Estimate & Std. Error & t value & Pr($>$$|$t$|$) \\ 
  \hline
  (Intercept) & 55.6979 & 2.0335 & 27.3898 & 0.0000 \\
  \texttt{td} & -3.5892 & 3.8915 & -0.9223 & 0.3571 \\
  \texttt{exacer} & 0.8697 & 3.2249 & 0.2697 & 0.7876 \\
  \texttt{panss} & 0.6213 & 0.0581 & 10.7015 & 0.0000 \\
  \texttt{olan} & 3.2705 & 1.6885 & 1.9370 & 0.0054 \\
  \texttt{panss}*\texttt{olan} & -0.0136 & 0.0583 & -0.2326 & 0.8162 \\
   \hline
\end{tabular}
\caption{\label{Phase2TPan}
 Summary of the fitted coefficients for PANSS outcome, Phase 2
  Tolerability arm. $N = 295$.}
\end{center}
\end{table}

\begin{table}[H]
\begin{center}
\begin{tabular}{rrrrr}
  \hline
 & Estimate & Std. Error & t value & Pr($>$$|$t$|$) \\ 
  \hline
  (Intercept) & 47.7079 & 0.8138 & 58.6223 & 0.0000 \\
  \texttt{td} & 0.7497 & 1.5619 & 0.4800 & 0.6316 \\
  \texttt{exacer} & -0.9157 & 1.2952 & -0.7070 & 0.4801 \\
  \texttt{bmi} & 0.9166 & 0.0228 & 40.1616 & 0.0000 \\
  \texttt{olan} & -3.9836 & 0.6708 & -5.9383 & 0.0000 \\
  \texttt{bmi}*\texttt{olan} & -0.0124 & 0.0228 & -0.5464 & 0.5852 \\
   \hline
\end{tabular}
\caption{\label{Phase2TBmi}
 Summary of the fitted coefficients for BMI outcome, Phase 2
  Tolerability arm. $N = 295$.}
\end{center}
\end{table}

Tables \ref{Phase2TPan} and \ref{Phase2TBmi} show the models estimated
from the Phase 2 tolerability data. As expected, olanzapine appears to
be benificial if one considers the PANSS ($Y$) outcome, but
detrimental if one considers the BMI ($Z$) outcome. This is borne out
in Figure~\ref{fig:CATIEPhase2Tol}, where we see that the predictions
of $(r_Y,r_Z)$ for all of the patient histories in our dataset fall in the
lower-right region of the plot, where both treatments are recommended
because they conflict with each other according to the two
outcomes. 

\renewcommand{\Dy}{5}
\renewcommand{\Dz}{5}
\begin{figure}
\begin{center}
\begin{tikzpicture}[y=0.1cm,x=0.1cm,scale=4]

%Central region of indifference
%\pattern[pattern=dots,pattern color=lightgray] (-\Dy,-\Dz) rectangle (\Dy,\Dz);
%Lower-right
%\pattern[pattern=dots,pattern color=lightgray] (\Dy,-\Dz) rectangle (1,-1);
%Upper-left
%\pattern[pattern=dots,pattern color=lightgray] (-\Dy,\Dz) rectangle (-1,1);

\pgfmathparse{-2*\Dy}
\let\LeftEnd\pgfmathresult
\pgfmathparse{3*\Dy}
\let\RightEnd\pgfmathresult
\pgfmathparse{-2*\Dz}
\let\BottomEnd\pgfmathresult
\pgfmathparse{2*\Dz}
\let\TopEnd\pgfmathresult

%Prefer treatment 1
\pattern[pattern=vertical lines,pattern color=lightgray] (-\Dy,\Dz) rectangle (\Dy,\TopEnd);
\pattern[pattern=vertical lines,pattern color=lightgray] (\Dy,\Dz) rectangle (\RightEnd,\TopEnd);
\pattern[pattern=vertical lines,pattern color=lightgray] (\Dy,-\Dz) rectangle (\RightEnd,\Dz);

%Prefer treatment -1
\pattern[pattern=horizontal lines,pattern color=lightgray] (-\Dy,\Dz) rectangle (\LeftEnd,-\Dz);
\pattern[pattern=horizontal lines,pattern color=lightgray] (-\Dy,-\Dz) rectangle (\LeftEnd,\BottomEnd);
\pattern[pattern=horizontal lines,pattern color=lightgray] (\Dy,-\Dz) rectangle (-\Dy,\BottomEnd);

%Thick lines
\newcommand{\lnth}{2pt}
\draw[line width=\lnth] (-\Dy,-\Dz) rectangle (\Dy,\Dz);
\draw[line width=\lnth] (-\Dy,\Dz) -- (-\Dy,\TopEnd);
\draw[line width=\lnth] (-\Dy,\Dz) -- (\LeftEnd,\Dz);
\draw[line width=\lnth] (\Dy,-\Dz) -- (\Dy,\BottomEnd);
\draw[line width=\lnth] (\Dy,-\Dz) -- (\RightEnd,-\Dz);

%Y direction (horizontal)
\draw (\LeftEnd,0) -- (\RightEnd,0);

%Delta Z bounds
\draw[dotted] (\LeftEnd,\Dz) -- (\RightEnd,\Dz);
\draw[dotted] (\LeftEnd,-\Dz) -- (\RightEnd,-\Dz);

%Z direction (vertical)
\draw (0,\BottomEnd) -- (0,\TopEnd);

%Delta Y bounds
\draw[dotted] (\Dy,\BottomEnd) -- (\Dy,\TopEnd);
\draw[dotted] (-\Dy,\BottomEnd) -- (-\Dy,\TopEnd);

%Treatment labels
\node at (\LeftEnd,\TopEnd) [fill=white,anchor=north west,outer sep=0pt] 
   {\parbox{2.5em}{{\centering Output:\\$\{-1,1\}$ \\ }}};
\node at (0,0) [fill=white] {Output: $\{-1,1\}$};
\node at (\RightEnd,-\Dz) [fill=white,anchor=north east,outer sep=10pt] 
   {\parbox{4em}{{\centering Output:\\$\{-1,1\}$ \\ }}};

\node at (-\Dy,-\Dz) [fill=white,anchor=north east,outer sep=5pt]
   {\parbox{3.25em}{{\centering Output:\\$\{-1\}$ \\ }}};
\node at (\RightEnd,\TopEnd) [fill=white,anchor=north east,outer sep=20pt] {Output: $\{1\}$};%

%Axis labels
\node at (0,\TopEnd) [anchor=south] {$r_Z(h)$};
\node at (\RightEnd,0) [anchor=west] {$r_Y(h)$};

%Delta Y ticks
\node at (\Dy,\TopEnd) [anchor=south] {$\Delta_Y$};
\node at (-\Dy,\TopEnd) [anchor=south] {$-\Delta_Y$};

%Delta Z ticks
\node at (\RightEnd,\Dz) [anchor=west] {$\Delta_Z$};
\node at (\RightEnd,-\Dz) [anchor=west] {$-\Delta_Z$};

\draw plot[only marks,mark=*,mark options={scale=0.05}] file {catie_phase2Tdata.txt};

\end{tikzpicture}

\end{center}
\caption{\label{fig:CATIEPhase2Tol}Diagram showing how the output of $\pi^\ideal(h)$ depends on
  $\Delta_Y$ and $\Delta_Z$, and on the location of the point
  $(r_Y(h),r_Z(h))$, for the Phase 2 Tolerability arm.}
\end{figure}
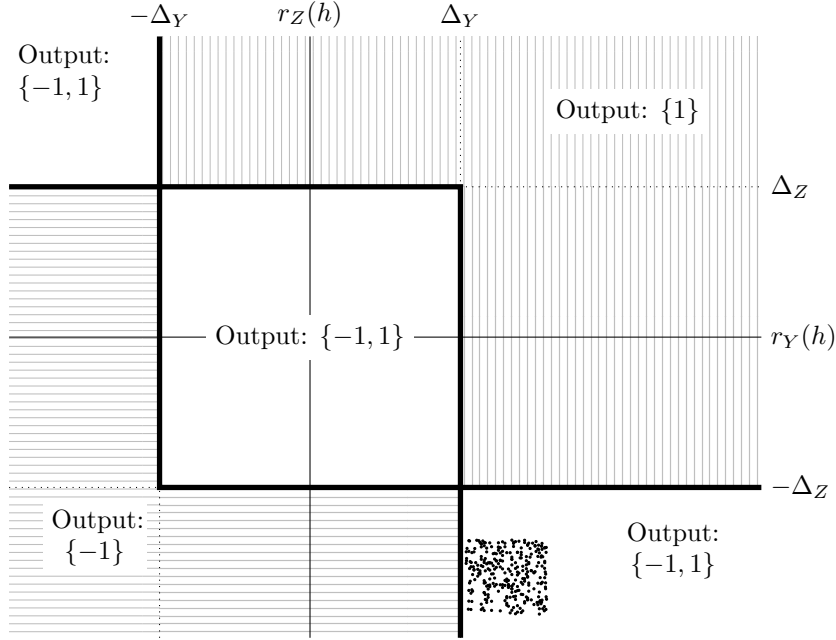

\subsubsection{Phase 2 Efficacy}

%%%%%%% EFFICACY

\begin{table}[H]
\begin{center}
\begin{tabular}{rrrrr}
  \hline
 & Estimate & Std. Error & t value & Pr($>$$|$t$|$) \\ 
  \hline
  (Intercept) & 54.7307 & 4.4225 & 12.3756 & 0.0000 \\
  \texttt{td} &  1.1844 & 7.9962 & 0.1481 & 0.8828 \\
  \texttt{exacer} & -3.0871 & 6.9329 & -0.4453 & 0.6580 \\
  \texttt{panss} & 0.6363 & 0.1299 & 4.898 & 0.0000 \\
  \texttt{cloz} &  9.2920 & 3.7722 & 2.463 & 0.0173 \\
  \texttt{panss}*\texttt{cloz} & 0.0220 & 0.1312 & 0.1673 & 0.8678 \\
   \hline
\end{tabular}
\caption{\label{Phase2EPan}
 Summary of the fitted coefficients for PANSS outcome, Phase 2
  Efficacy arm.}
\end{center}
\end{table}

\begin{table}[H]
\begin{center}
\begin{tabular}{rrrrr}
  \hline
 & Estimate & Std. Error & t value & Pr($>$$|$t$|$) \\ 
  \hline
  (Intercept) & 50.7367 & 1.7384 & 29.1863 & 0.0000 \\
  \texttt{td} & -5.2649 & 3.0542 & -1.7238 & 0.0916 \\
  \texttt{exacer} & -2.1386 & 2.8634 & -0.7469 & 0.4586 \\
  \texttt{bmi} & 0.9277 & 0.0507 & 18.2857 & 0.0000 \\
  \texttt{cloz} & -1.1109 & 1.3582 & -0.8179 & 0.4173 \\
  \texttt{bmi}*\texttt{cloz} & -0.0592 & 0.0525 & -1.1272 & 0.2650 \\
   \hline
\end{tabular}
\caption{\label{Phase2EBmi}
 Summary of the fitted coefficients for BMI outcome, Phase 2
  Efficacy arm. $N = 56$}
\end{center}
\end{table}

Tables \ref{Phase2EPan} and \ref{Phase2EBmi} show the models estimated
from the Phase 2 efficacy data. As expected, clozapine appears to
be beneficial if one considers the PANSS ($Y$) outcome. Furthermore,
there is weak evidence that clozapine is detrimental if one considers
the BMI ($Z$) outcome. This is borne out in
Figure~\ref{fig:CATIEPhase2Eff}, where we see that the predictions of
$(r_Y,r_Z)$ for all of the patient histories in our dataset are to the right of
$r_Y = \Delta_Y$, indicating that clozapine is predicted to be the
better choice for all patients in the dataset when considering only the
PANSS outcome. Furthermore, for most of these, clozapine is not
significantly worse than the other (aggregate) treatment according to
BMI; thus for most of the histories only clozapine (i.e.\ $\{1\}$ would
be recommended. We found that for patients with a BMI covariate
greater than about 24 (i.e.\ those among the top best 25 percent
according to BMI\footnote{Recall the negative coding (higher
  percentiles are better) and the shift by 50: It is the patients
  whose BMI is better than the 74th percentile who are recommended
  both treatments $\{-1,1\}$.}), however, clozapine is predicted to
perform clinically significantly worse according to the BMI outcome,
and both treatments (i.e. $\{-1,1\}$) would be recommended for these
patients.

\renewcommand{\Dy}{5}
\renewcommand{\Dz}{5}
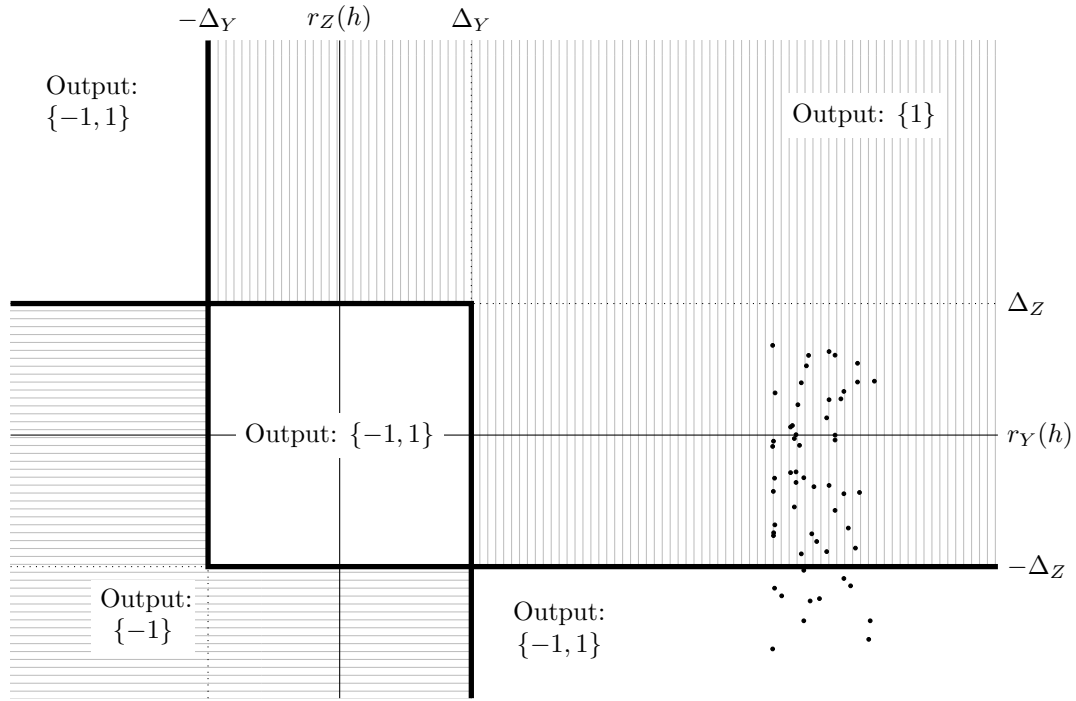
\begin{figure}
\begin{center}
\begin{tikzpicture}[y=0.1cm,x=0.1cm,scale=3.5]

%Central region of indifference
%\pattern[pattern=dots,pattern color=lightgray] (-\Dy,-\Dz) rectangle (\Dy,\Dz);
%Lower-right
%\pattern[pattern=dots,pattern color=lightgray] (\Dy,-\Dz) rectangle (1,-1);
%Upper-left
%\pattern[pattern=dots,pattern color=lightgray] (-\Dy,\Dz) rectangle (-1,1);

\pgfmathparse{-2.5*\Dy}
\let\LeftEnd\pgfmathresult
\pgfmathparse{5*\Dy}
\let\RightEnd\pgfmathresult
\pgfmathparse{-2*\Dz}
\let\BottomEnd\pgfmathresult
\pgfmathparse{3*\Dz}
\let\TopEnd\pgfmathresult

%Prefer treatment 1
\pattern[pattern=vertical lines,pattern color=lightgray] (-\Dy,\Dz) rectangle (\Dy,\TopEnd);
\pattern[pattern=vertical lines,pattern color=lightgray] (\Dy,\Dz) rectangle (\RightEnd,\TopEnd);
\pattern[pattern=vertical lines,pattern color=lightgray] (\Dy,-\Dz) rectangle (\RightEnd,\Dz);

%Prefer treatment -1
\pattern[pattern=horizontal lines,pattern color=lightgray] (-\Dy,\Dz) rectangle (\LeftEnd,-\Dz);
\pattern[pattern=horizontal lines,pattern color=lightgray] (-\Dy,-\Dz) rectangle (\LeftEnd,\BottomEnd);
\pattern[pattern=horizontal lines,pattern color=lightgray] (\Dy,-\Dz) rectangle (-\Dy,\BottomEnd);

%Thick lines
\newcommand{\lnth}{2pt}
\draw[line width=\lnth] (-\Dy,-\Dz) rectangle (\Dy,\Dz);
\draw[line width=\lnth] (-\Dy,\Dz) -- (-\Dy,\TopEnd);
\draw[line width=\lnth] (-\Dy,\Dz) -- (\LeftEnd,\Dz);
\draw[line width=\lnth] (\Dy,-\Dz) -- (\Dy,\BottomEnd);
\draw[line width=\lnth] (\Dy,-\Dz) -- (\RightEnd,-\Dz);

%Y direction (horizontal)
\draw (\LeftEnd,0) -- (\RightEnd,0);

%Delta Z bounds
\draw[dotted] (\LeftEnd,\Dz) -- (\RightEnd,\Dz);
\draw[dotted] (\LeftEnd,-\Dz) -- (\RightEnd,-\Dz);

%Z direction (vertical)
\draw (0,\BottomEnd) -- (0,\TopEnd);

%Delta Y bounds
\draw[dotted] (\Dy,\BottomEnd) -- (\Dy,\TopEnd);
\draw[dotted] (-\Dy,\BottomEnd) -- (-\Dy,\TopEnd);

%Treatment labels
\node at (\LeftEnd,\TopEnd) [fill=white,anchor=north west,outer sep=10pt] 
   {\parbox{2.5em}{{\centering Output:\\$\{-1,1\}$ \\ }}};
\node at (0,0) [fill=white] {Output: $\{-1,1\}$};
\node at (\Dy,-\Dz) [fill=white,anchor=north west,outer sep=10pt] 
   {\parbox{4em}{{\centering Output:\\$\{-1,1\}$ \\ }}};
\node at (-\Dy,-\Dz) [fill=white,anchor=north east,outer sep=5pt]
   {\parbox{3.25em}{{\centering Output:\\$\{-1\}$ \\ }}};
\node at (\RightEnd,\TopEnd) [fill=white,anchor=north east,outer sep=20pt] {Output: $\{1\}$};%

%Axis labels
\node at (0,\TopEnd) [anchor=south] {$r_Z(h)$};
\node at (\RightEnd,0) [anchor=west] {$r_Y(h)$};

%Delta Y ticks
\node at (\Dy,\TopEnd) [anchor=south] {$\Delta_Y$};
\node at (-\Dy,\TopEnd) [anchor=south] {$-\Delta_Y$};

%Delta Z ticks
\node at (\RightEnd,\Dz) [anchor=west] {$\Delta_Z$};
\node at (\RightEnd,-\Dz) [anchor=west] {$-\Delta_Z$};

\draw plot[only marks,mark=*,mark options={scale=0.1}] file {catie_phase2Edata.txt};

\end{tikzpicture}

\end{center}
\caption{\label{fig:CATIEPhase2Eff}Diagram showing how the output of $\pi^\ideal(h)$ depends on
  $\Delta_Y$ and $\Delta_Z$, and on the location of the point
  $(r_Y(h),r_Z(h))$, for the Phase 2 Efficacy arm. $N = 56$.}
\end{figure}

\subsubsection{Phase 1}

%%% PHASE 1

\begin{table}[H]
\begin{center}
\begin{tabular}{rrrrr}
  \hline
 & Estimate & Std. Error & t value & Pr($>$$|$t$|$) \\ 
  \hline
(Intercept) & 57.3576 & 1.0382 & 55.2486 & 0.0000 \\
  \texttt{td} & -5.8840 & 2.0415 & -2.8823 & 0.0004 \\
  \texttt{exacer} & 1.1234 & 1.6471 & 0.6821 & 0.4954 \\
  \texttt{panss} & 0.5332 & 0.0318 & 16.7574 & 0.0000 \\
  \texttt{perp} & -2.6691 & 0.9505 & -2.8081 & 0.0051 \\
  \texttt{panss}*\texttt{perp} & 0.0778 & 0.0317 & 2.4531 & 0.0143 \\
   \hline
\end{tabular}
\caption{\label{Phase1Pan}
  Example summary of the fitted coefficients for PANSS outcome,
  Phase 1, based on a randomly chosen feasible decision rule for Phase
  2. $N = 974$}
\end{center}
\end{table}

\begin{table}[H]
\begin{center}
\begin{tabular}{rrrrr}
  \hline
 & Estimate & Std. Error & t value & Pr($>$$|$t$|$) \\ 
  \hline
  (Intercept) & 49.0622 & 0.5084 & 96.5089 & 0.0000 \\
  \texttt{td} & 1.2004 & 1.0069 & 1.1922 & 0.2335 \\
  \texttt{exacer} & -2.7812 & 0.8175 & -3.4022 & 0.0007 \\
  \texttt{bmi} & 0.9134 & 0.0163 & 55.2402 & 0.0000 \\
  \texttt{perp} & 1.8266 & 0.4659 & 3.9197 & 0.0001 \\
  \texttt{bmi}*\texttt{perp} & -0.0250 & 0.0163 & -1.5388 & 0.1242 \\
   \hline
\end{tabular}
\caption{\label{Phase1Bmi}
  Example summary of the fitted coefficients for BMI outcome,
  based on a randomly chosen feasible decision rule for Phase 2. $N = 974$.}
\end{center}
\end{table}

We now consider Phase 1. Recall that given any history $h_1$ at Phase 1, our
predicted values $(r_Y,r_Z)$ for that history depend not only on the
history itself but on the future decision rule that will be followed
subsequently. For illustrative purposes, Tables \ref{Phase1Pan} and
\ref{Phase1Bmi} show the models estimated from the Phase 1 data
assuming a particular feasible decision rule for Phase 2 chosen from the
$61,659$ feasible Phase 2 decision rules enumerated by our algorithm. (The
estimated coefficients would be different had we used a different
Phase 2 decision rule.) For this particular future decision rule, perphenazine performs
somewhat worse according to PANSS than the atypical antipsychotics,
and somewhat better according to BMI.

Whereas for the Phase 2 analyses we showed plots of different
$(r_Y,r_Z)$ for different histories, for Phase 1, we will show different
$(r_Y,r_Z)$ for a \textit{fixed} history at Phase 1 as we vary the Phase
2 decision rule. Recall that our treatment recommendation for Phase 1 is the
union over all feasible future decision rules of the treatments
recommended for each future decision rule. Figure~\ref{fig:CATIEP1} shows the
possible values of $(r_Y,r_Z)$. From Figure~\ref{fig:CATIEP1}, we can
see that for some future decision rules only treatment $-1$ is recommended,
but for others the set $\{-1,1\}$ is recommended. Taking the union, we
recommend the set $\{-1,1\}$ for this history at Phase 1.

\renewcommand{\Dy}{5}
\renewcommand{\Dz}{5}
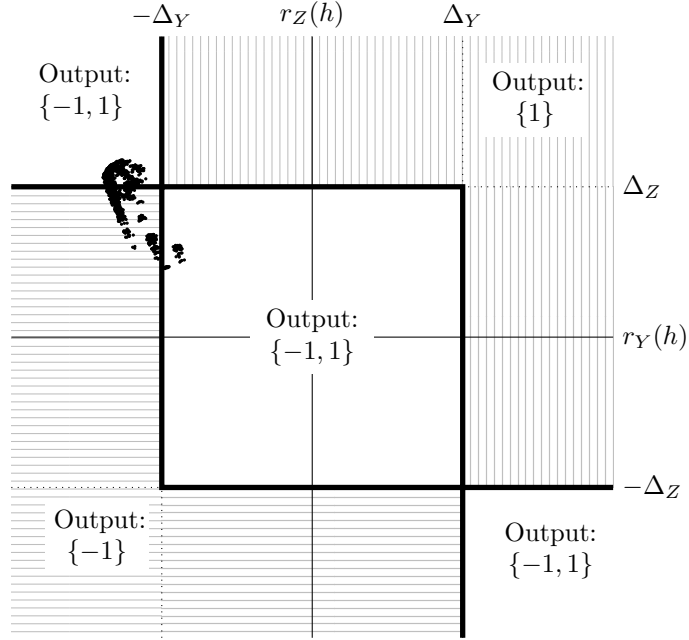
\begin{figure}
\begin{center}
\begin{tikzpicture}[y=0.1cm,x=0.1cm,scale=4]

%Central region of indifference
%\pattern[pattern=dots,pattern color=lightgray] (-\Dy,-\Dz) rectangle (\Dy,\Dz);
%Lower-right
%\pattern[pattern=dots,pattern color=lightgray] (\Dy,-\Dz) rectangle (1,-1);
%Upper-left
%\pattern[pattern=dots,pattern color=lightgray] (-\Dy,\Dz) rectangle (-1,1);

\pgfmathparse{-2*\Dy}
\let\LeftEnd\pgfmathresult
\pgfmathparse{2*\Dy}
\let\RightEnd\pgfmathresult
\pgfmathparse{-2*\Dz}
\let\BottomEnd\pgfmathresult
\pgfmathparse{2*\Dz}
\let\TopEnd\pgfmathresult

%Prefer treatment 1
\pattern[pattern=vertical lines,pattern color=lightgray] (-\Dy,\Dz) rectangle (\Dy,\TopEnd);
\pattern[pattern=vertical lines,pattern color=lightgray] (\Dy,\Dz) rectangle (\RightEnd,\TopEnd);
\pattern[pattern=vertical lines,pattern color=lightgray] (\Dy,-\Dz) rectangle (\RightEnd,\Dz);

%Prefer treatment -1
\pattern[pattern=horizontal lines,pattern color=lightgray] (-\Dy,\Dz) rectangle (\LeftEnd,-\Dz);
\pattern[pattern=horizontal lines,pattern color=lightgray] (-\Dy,-\Dz) rectangle (\LeftEnd,\BottomEnd);
\pattern[pattern=horizontal lines,pattern color=lightgray] (\Dy,-\Dz) rectangle (-\Dy,\BottomEnd);

%Thick lines
\newcommand{\lnth}{2pt}
\draw[line width=\lnth] (-\Dy,-\Dz) rectangle (\Dy,\Dz);
\draw[line width=\lnth] (-\Dy,\Dz) -- (-\Dy,\TopEnd);
\draw[line width=\lnth] (-\Dy,\Dz) -- (\LeftEnd,\Dz);
\draw[line width=\lnth] (\Dy,-\Dz) -- (\Dy,\BottomEnd);
\draw[line width=\lnth] (\Dy,-\Dz) -- (\RightEnd,-\Dz);

%Y direction (horizontal)
\draw (\LeftEnd,0) -- (\RightEnd,0);

%Delta Z bounds
\draw[dotted] (\LeftEnd,\Dz) -- (\RightEnd,\Dz);
\draw[dotted] (\LeftEnd,-\Dz) -- (\RightEnd,-\Dz);

%Z direction (vertical)
\draw (0,\BottomEnd) -- (0,\TopEnd);

%Delta Y bounds
\draw[dotted] (\Dy,\BottomEnd) -- (\Dy,\TopEnd);
\draw[dotted] (-\Dy,\BottomEnd) -- (-\Dy,\TopEnd);

%Treatment labels
\node at (\LeftEnd,\TopEnd) [fill=white,anchor=north west,outer sep=7pt] 
   {\parbox{2.5em}{{\centering Output:\\$\{-1,1\}$ \\ }}};
\node at (0,0) [fill=white]
   {\parbox{4em}{{\centering Output:\\$\{-1,1\}$ \\ }}};
\node at (\Dy,-\Dz) [fill=white,anchor=north west,outer sep=10pt] 
   {\parbox{4em}{{\centering Output:\\$\{-1,1\}$ \\ }}};
\node at (-\Dy,-\Dz) [fill=white,anchor=north east,outer sep=5pt]
   {\parbox{3.25em}{{\centering Output:\\$\{-1\}$ \\ }}};
\node at (\RightEnd,\TopEnd) [fill=white,anchor=north east,outer sep=10pt] 
   {\parbox{3.25em}{{\centering Output:\\$\{1\}$ \\ }}};

%Axis labels
\node at (0,\TopEnd) [anchor=south] {$r_Z(h)$};
\node at (\RightEnd,0) [anchor=west] {$r_Y(h)$};

%Delta Y ticks
\node at (\Dy,\TopEnd) [anchor=south] {$\Delta_Y$};
\node at (-\Dy,\TopEnd) [anchor=south] {$-\Delta_Y$};

%Delta Z ticks
\node at (\RightEnd,\Dz) [anchor=west] {$\Delta_Z$};
\node at (\RightEnd,-\Dz) [anchor=west] {$-\Delta_Z$};

\draw plot[only marks,mark=*,mark options={scale=0.05}] file
{catie_phase1data.txt};
%using file 49
%State: 265 PANNS=-2.547945205479452113e+01 BMI=-1.565068493150685924e+01
\end{tikzpicture}

\end{center}
\caption{\label{fig:CATIEP1}Diagram showing how the output of $\pi^\ideal(h)$ depends on
  $\Delta_Y$ and $\Delta_Z$, and on the location of the point
  $(r_Y(h),r_Z(h))$, for Phase 1, at history $\mathtt{panss} =
  -25.5$,$\mathtt{bmi} = -15.6$).}
\end{figure}

\section{Discussion} 
We proposed set-valued dynamic treatment regimes as a method for
adapting treatment recommendations to the evolving health status of a
patient in the presence of competing outcomes. Our proposed
methodology deals with the reality that there is typically no
universally good treatment for chronic illnesses like depression or
schizophrenia by identifying when a trade-off between effectiveness
and side-effects must be made.  Although computation of the set-valued
dynamic treatment regimes requires solving a difficult enumeration
problem, we offered an efficient algorithm that makes use of existing
linear mixed integer programming software packages. We demonstrated
the use of our method using data from one-stage and two-stage
randomized clinical trials.

Our proposed methodology avoids the construction of composite
outcomes, a process which may be undesirable: constructing a composite
outcome requires combining outcomes that are on different scales, the
`optimal trade-off' between two (or more) outcomes is likely to be
patient-specific, and the assumption that a linear trade-off is
sufficient to describe all possible patient preferences may be
unrealistic.

%Patients will
%not fully understand how they would ideally balance competing outcomes
%until treatment has already started and they begin to experience these
%outcomes first-hand.

%%%Dan: I think if we want to talk about this we have to like really
%%%have an in-depth discussion and I don't know if it's worth
%%%it... Maybe it is. I dunno.

There are a number of directions in which this work can be extended.
The appendix provides an extension to the case with two decision
points but an arbitrary number of treatment choices available at each
stage. Interestingly, our enumeration problem is closely related to
\textit{transductive learning}, a classification problem setting where
only a subset of the available training data is labeled, and the task
is to predict labels at the unlabeled points in the training data.  By
finding a minimum-norm solution for $\psi$ subject to our constraints,
we could produce the transductive labeling that induces the maximum
margin linear separator. In essence, our algorithm would then
correspond to a linear separable transductive support vector machine
(SVM) ~\citep{CortesSVM}. This observation leads to a possible
criterion for evaluating feasible decision rules: we hypothesize that
the greater the induced margin, the more ``intuitive'' the decision
rule, because large-margin decision rules avoid giving very similar
patients different treatments. If the number of feasible future
decision rules becomes impractically large, we may wish to keep only
the most ``separable'' ones when computing the union at the first
stage.  We are currently pursuing this line of research.

\section*{Acknowledgements}

Data used in the preparation of this article were obtained from the
limited access datasets distributed from the NIH-supported ``Clinical
Antipsychotic Trials of Intervention Effectiveness in Schizophrenia''
(CATIE-Sz). This is a multisite, clinical trial of persons with
schizophrenia comparing the effectiveness of randomly assigned
medication treatment. The study was supported by NIMH Contract
\#N01MH90001 to the University of North Carolina at Chapel Hill. The
ClinicalTrials.gov identifier is NCT00014001.  

This manuscript reflects the views of the authors and may not reflect
the opinions or views of the CATIE-Sz Study Investigators or the NIH.

\bibliography{setValuedCites}
\bibliographystyle{plainnat}

\appendix
\section{Multiple Treatments}

To develop our method for the binary treatment case, we considered
working models of the form $Q_t(h_t, a_t) = h_{t,1}^{\T}\beta_{t} +
a_t h_{t,2}^{\T}\psi_{t}$, and we defined our estimated optimal treatment
for history $h_t$ to be $\hat{\pi}_t (h_t) = \arg\max_{a_{t} \in
  \{-1,1\}} \hat{Q}(h_t,a_t)$. If there are more than two levels of
treatment (suppose $a_t$ belongs to a discrete set $\mathcal{A}_t$) we
define estimated working models of the form
\begin{equation}
\hat{Q}_t(h_t, a_t) = h_{t,1}^{\T}\hat\beta_{t} + \phi(a_t,h_{t,2})^{\T}\hat\psi_{t}
\end{equation}
where the vector-valued function $\phi$ describes an arbitrary
encoding of $a_t$ along with any desired interactions between the
encoding and $h_{t,2}$. For example, if $\mathcal{A}_t =  \{1,2,3\}$ and
$h_{t,2}$ is a scalar, we might
use
\begin{equation}
\phi(a_t,h_{t,2}) = 
\begin{cases}
(0, 0, 0, 0)^\T & \text{if $a_t = 1$} \\
(1, 0, h_{t,2}, 0)^\T & \text{if $a_t = 2$} \\
(0, 1, 0, h_{t,2})^\T & \text{if $a_t = 3$}
\end{cases}
\end{equation}
to produce a model that incorporates a main effect of treatment as
well as an interaction between treatment and $h_{t,2}$. The choice of
how to encode factors and interactions has been well-studied
~\citep{wu2000experiments}. Regardless of the specific choice of encoding,
our estimated DTR is $\hat{\pi}_t (h_t) = \arg\max_{a_{t} \in
  \mathcal{A}_t} \hat{Q}(h_t,a_t)$.

\subsection{Producing the Set-Valued Decision Rule}\label{ap:setvaluedrule}

In order to identify the set of treatments that should be recommended
for a particular $h_t$, we consider all pairs of treatments and
identify those which are never eliminated in any pairwise comparison
according to our definition of `clincial significance.' Continuing our
example, if for a particular $h_t$ we find that considering $1$ and
$2$ recommends the set $\{1,2\}$ and that considering $1$ and $3$
recommends the set $\{1\}$, then we would include treatment $1$ in our
recommended set for $h_t$. Note that we can also infer that $3$ would
\textit{not} be recommended, since in a pairwise comparison with $1$ it is
eliminated.

\newcommand{\Aset}{\ensuremath{\mathcal{A}}}

\subsection{Enumerating the Feasible Decision Rules}\label{ap:feasible}
To construct the MIP describing the feasible decision rules, we introduce $n \times |\Aset_t|$ indicator
variables $\alpha_{i,j}$ that indicate whether $\hat{\pi}(h^{(i)}_t) = j$ or not. We then
impose the following constraints:
\begin{gather}
\forall i \in 1...n,~j\in 1...|\Aset_t|,~\alpha_{i,j} \in \{0,1\} \label{eq:binaryconst}\\
\forall i \in 1...n,~\sum_j \alpha_{i,j} =
1 \label{eq:exactlyoneconst} \\
\forall i \in 1...n,~\forall j \in 1...|\Aset_t|,~\alpha_{i,j} = 1 \implies
  \forall k \ne j,~ (\phi(h_t^{(i)},j) - \phi(h_t^{(i)},k))^\T \psi_2 \ge 1. \label{eq:marginconst}
\end{gather}
Constraints (\ref{eq:binaryconst}) ensure that the indicator variables
for the actions are binary. Constraints (\ref{eq:exactlyoneconst})
ensure that, for each example in our dataset, exactly one action
indicator variable is on. The indicator constraints in
(\ref{eq:marginconst}) ensure that if the indicator for action $j$
is on for the $i$th example, then weights must satisfy $j = \arg\max_a
\phi(s^i,a)^\T w$. Note that the margin condition (i.e. having the
constraint be $\ge 1$ rather than $\ge 0$) avoids a degenerate
solution with $\psi_2 = \mathbf{0}$.

The above constraints ensure that the $\alpha_{i,j}$ define a
treatment rule that can be represented as an $\arg\max$ in the given
covariate space. Imposing the additional constraint that the treatment
rule defined is compatible with a given set-valued treatment rule
$\tilde\pi$ is now trivial:
\begin{gather}
\forall i \in 1...n,~ \sum_{j \in \tilde\pi(h^{(i)})} \alpha_{i,j} = 1. \label{eq:compatible}
\end{gather}
Constraints (\ref{eq:compatible}) ensure that the indicator that turns
on for the $i$th example in the data must be one that indicates an
action already present in $\tilde\pi(h^{(i)})$. 

Using the approaches developed in Sections~\ref{ap:setvaluedrule} and
\ref{ap:feasible}, an estimation procedure analogous to that
described in Section~\ref{ss:twopoint} immediately follows.

\clearpage

\section{Regression Diagnostics}
\subsection{Nefazodone study}
\subsubsection*{Depression}
\begin{verbatim}
Residuals:
     Min       1Q   Median       3Q      Max 
-2.25326 -0.35043  0.02561  0.46046  1.47199 

Coefficients:
               Estimate Std. Error t value Pr(>|t|)    
(Intercept)   0.2867221  0.2157694   1.329   0.1848    
HAMDTOT       0.0324713  0.0073585   4.413 1.38e-05 ***
ROLFUN       -0.0009138  0.0009052  -1.009   0.3135    
RAND          0.1133281  0.2157694   0.525   0.5998    
ROLFUN:RAND   0.0018196  0.0009052   2.010   0.0452 *  
HAMDTOT:RAND  0.0010579  0.0073585   0.144   0.8858    
---
Signif. codes:  0 Ô***Õ 0.001 Ô**Õ 0.01 Ô*Õ 0.05 Ô.Õ 0.1 Ô Õ 1 

Residual standard error: 0.6544 on 330 degrees of freedom
Multiple R-squared: 0.1969,	Adjusted R-squared: 0.1847 
F-statistic: 16.18 on 5 and 330 DF,  p-value: 2.835e-14 
\end{verbatim}

\begin{figure}[H] %  figure placement: here, top, bottom, or page
   \centering
   \includegraphics[width=5in]{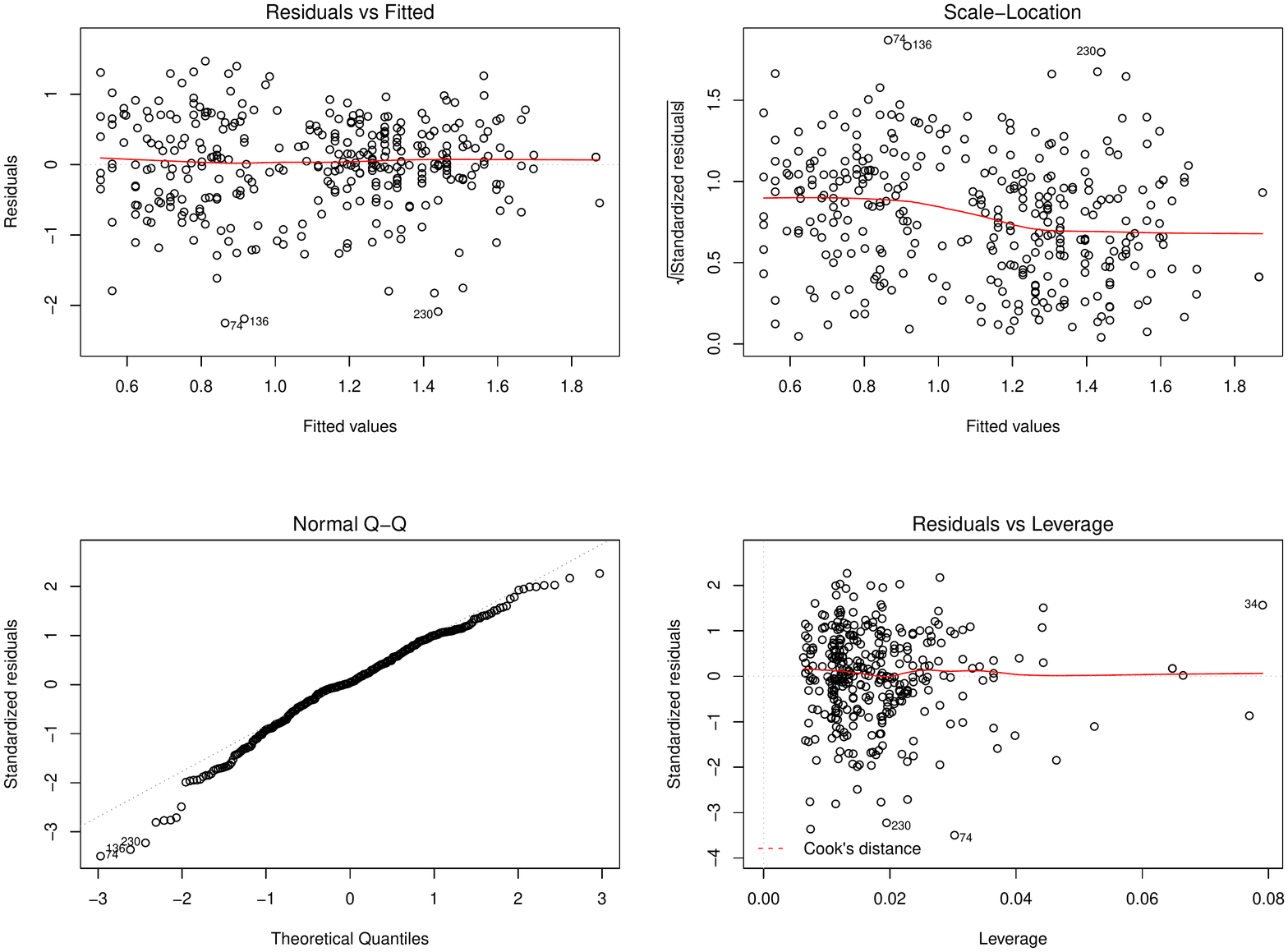}
   \label{fig:example}
\end{figure}

\subsubsection*{Physical Functioning}
\begin{verbatim}
Residuals:
    Min      1Q  Median      3Q     Max 
-43.741  -5.730   1.100   6.801  48.197 

Coefficients:
            Estimate Std. Error t value Pr(>|t|)    
(Intercept) 35.66068    4.83187   7.380 1.34e-12 ***
GENDER2     -3.44352    1.58505  -2.172 0.030543 *  
SLPSC1      -0.09793    0.34809  -0.281 0.778627    
PHYFUN       0.61981    0.04425  14.009  < 2e-16 ***
GENHEL       0.13838    0.04132   3.349 0.000906 ***
ROLFUN      -0.04361    0.02059  -2.118 0.034945 *  
MD_AGE      -0.12357    0.05826  -2.121 0.034682 *  
DYST_YES    -3.86101    1.48587  -2.598 0.009792 ** 
RAND        11.38947    3.82004   2.982 0.003086 ** 
SLPSC1:RAND -0.87366    0.34430  -2.538 0.011633 *  
PHYFUN:RAND -0.07143    0.03648  -1.958 0.051116 .  
---
Signif. codes:  0 Ô***Õ 0.001 Ô**Õ 0.01 Ô*Õ 0.05 Ô.Õ 0.1 Ô Õ 1 

Residual standard error: 13.22 on 324 degrees of freedom
  (1 observation deleted due to missingness)
Multiple R-squared: 0.5556,	Adjusted R-squared: 0.5419 
F-statistic: 40.51 on 10 and 324 DF,  p-value: < 2.2e-16 

\end{verbatim}

\begin{figure}[H] %  figure placement: here, top, bottom, or page
   \centering
   \includegraphics[width=5in]{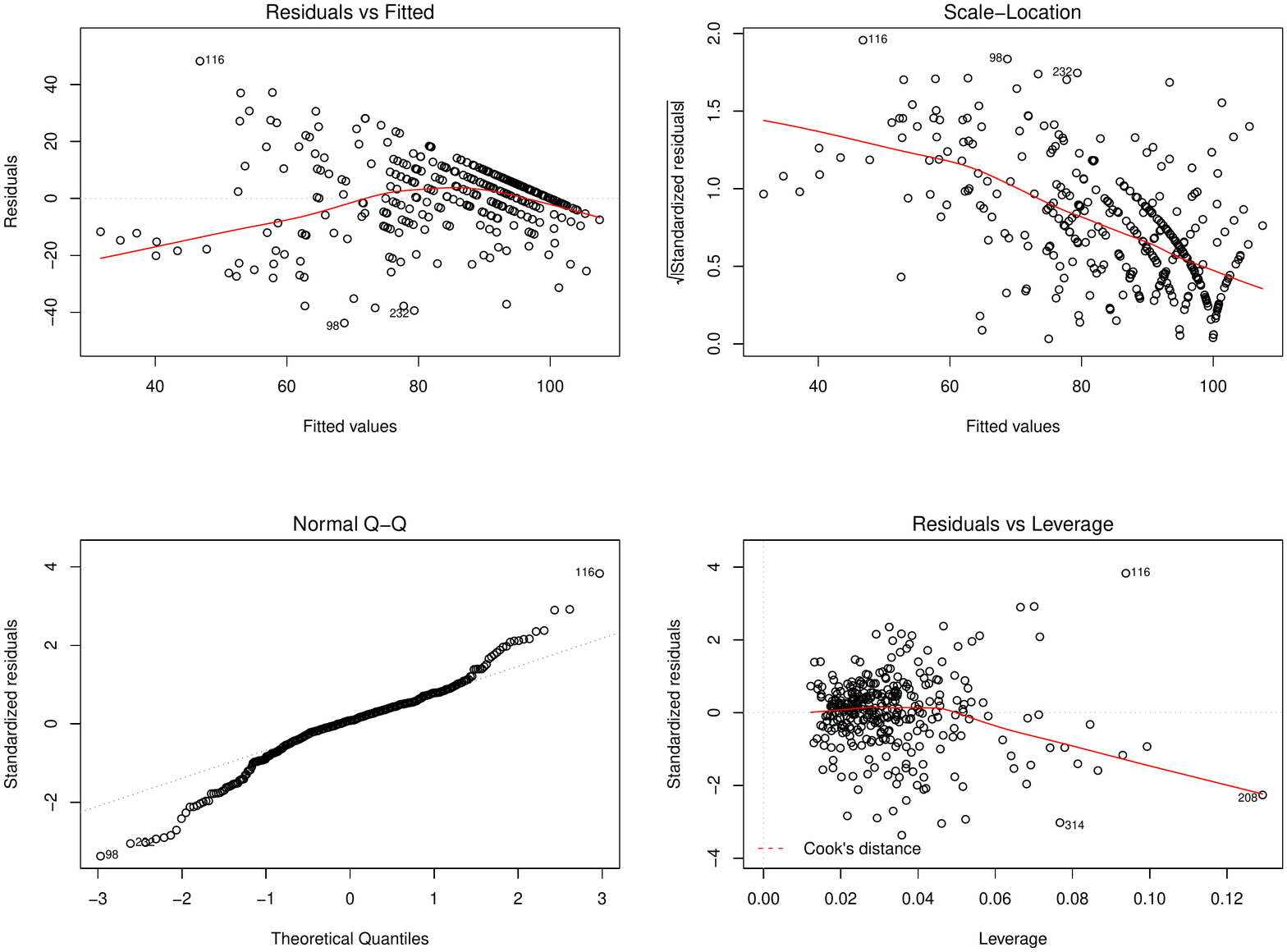}
   \label{fig:example}
\end{figure}

\subsection{CATIE}
\textbf{Note} that at Phase 1 the regression
estimators are non-regular, and that inference in this setting
requires additional care as many standard methods are not valid
~\citep{Laberetal}. Nonetheless we include the following
standard regression diagnostics to give a sense of model fit.

\newcommand{\diagfigwidth}{0.9}

\clearpage
\singlespace
\subsubsection{Phase 2 Tolerability: PANSS}

\begin{verbatim}
Residuals:
    Min      1Q  Median      3Q     Max 
-73.323 -17.868  -0.176  17.801  72.125 

Coefficients:
            Estimate Std. Error t value Pr(>|t|)    
(Intercept) 55.69794    2.03353  27.390   <2e-16 ***
TD          -3.58917    3.89155  -0.922   0.3571    
EXACER       0.86974    3.22493   0.270   0.7876    
PANSS        0.62132    0.05806  10.702   <2e-16 ***
OLAN         3.27049    1.68846   1.937   0.0537 .  
PANSS:OLAN  -0.01356    0.05829  -0.233   0.8162    
---
Signif. codes:  0 ‘***’ 0.001 ‘**’ 0.01 ‘*’ 0.05 ‘.’ 0.1 ‘ ’ 1 

Residual standard error: 24.47 on 289 degrees of freedom
Multiple R-squared: 0.3613,	Adjusted R-squared: 0.3503 
F-statistic:  32.7 on 5 and 289 DF,  p-value: < 2.2e-16 
\end{verbatim}
\includegraphics[width=\diagfigwidth\textwidth]{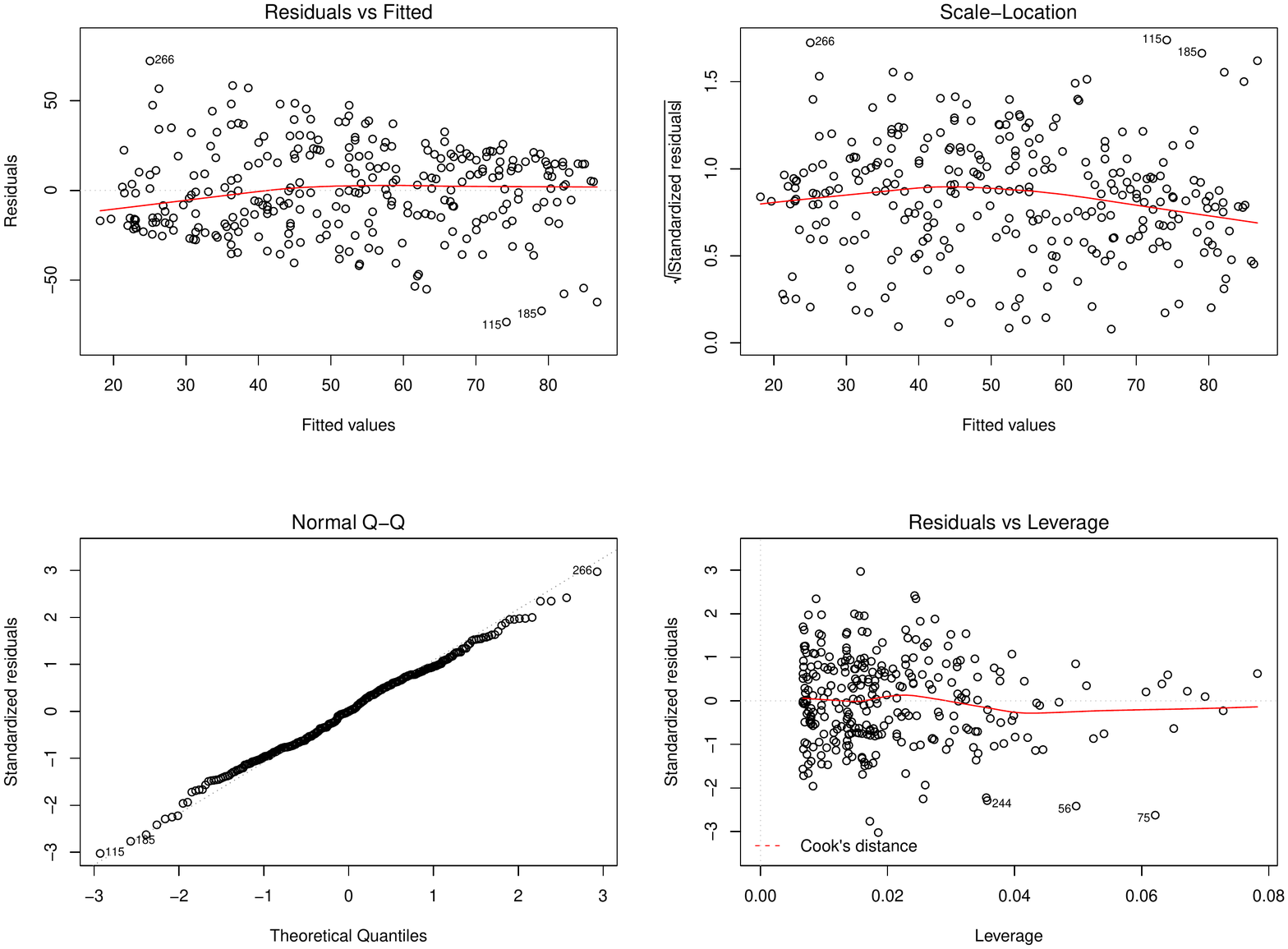}

\subsubsection{Phase 2 Tolerability: BMI}
\begin{verbatim}
Residuals:
    Min      1Q  Median      3Q     Max 
-38.343  -4.787  -0.097   4.886  54.400 

Coefficients:
            Estimate Std. Error t value Pr(>|t|)    
(Intercept) 47.70789    0.81382  58.622  < 2e-16 ***
TD           0.74972    1.56186   0.480    0.632    
EXACER      -0.91572    1.29519  -0.707    0.480    
BMI          0.91663    0.02282  40.162  < 2e-16 ***
OLAN        -3.98360    0.67083  -5.938 8.25e-09 ***
BMI:OLAN    -0.01245    0.02278  -0.546    0.585    
---
Signif. codes:  0 ‘***’ 0.001 ‘**’ 0.01 ‘*’ 0.05 ‘.’ 0.1 ‘ ’ 1 

Residual standard error: 9.822 on 289 degrees of freedom
Multiple R-squared: 0.8765,	Adjusted R-squared: 0.8744 
F-statistic: 410.4 on 5 and 289 DF,  p-value: < 2.2e-16 
\end{verbatim}
\includegraphics[width=\diagfigwidth\textwidth]{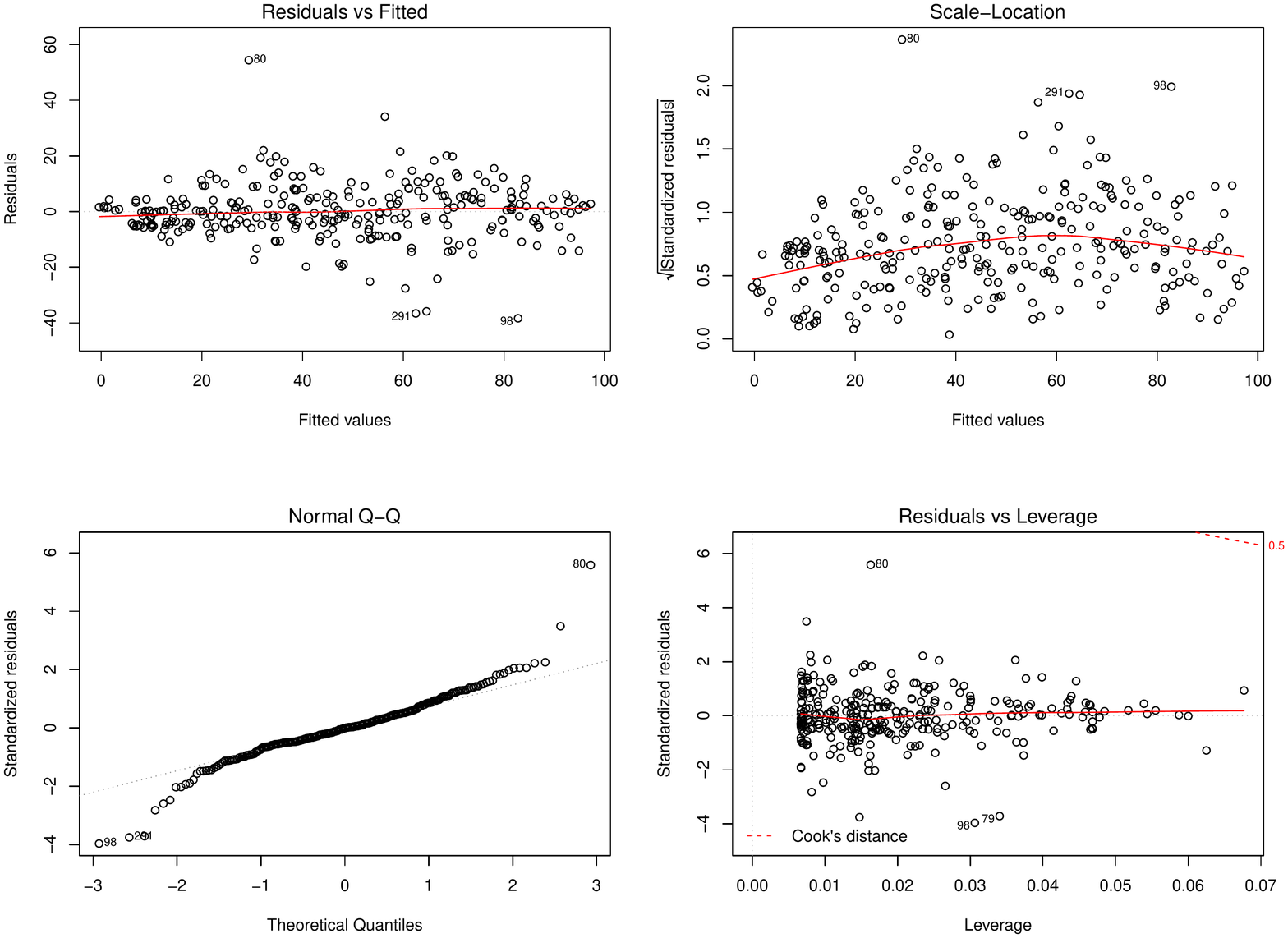}

\subsubsection{Phase 2 Efficacy: PANSS}
\begin{verbatim}
Residuals:
    Min      1Q  Median      3Q     Max 
-56.634 -15.947  -3.404  13.906  55.476 

Coefficients:
            Estimate Std. Error t value Pr(>|t|)    
(Intercept) 54.73068    4.42248  12.376  < 2e-16 ***
TD           1.18442    7.99619   0.148   0.8828    
EXACER      -3.08713    6.93291  -0.445   0.6580    
PANSS        0.63633    0.12991   4.898 1.06e-05 ***
CLOZ         9.29195    3.77220   2.463   0.0173 *  
PANSS:CLOZ   0.02196    0.13122   0.167   0.8678    
---
Signif. codes:  0 ‘***’ 0.001 ‘**’ 0.01 ‘*’ 0.05 ‘.’ 0.1 ‘ ’ 1 

Residual standard error: 23.25 on 50 degrees of freedom
Multiple R-squared: 0.3782,	Adjusted R-squared: 0.316 
F-statistic: 6.082 on 5 and 50 DF,  p-value: 0.0001793
\end{verbatim}
\includegraphics[width=\diagfigwidth\textwidth]{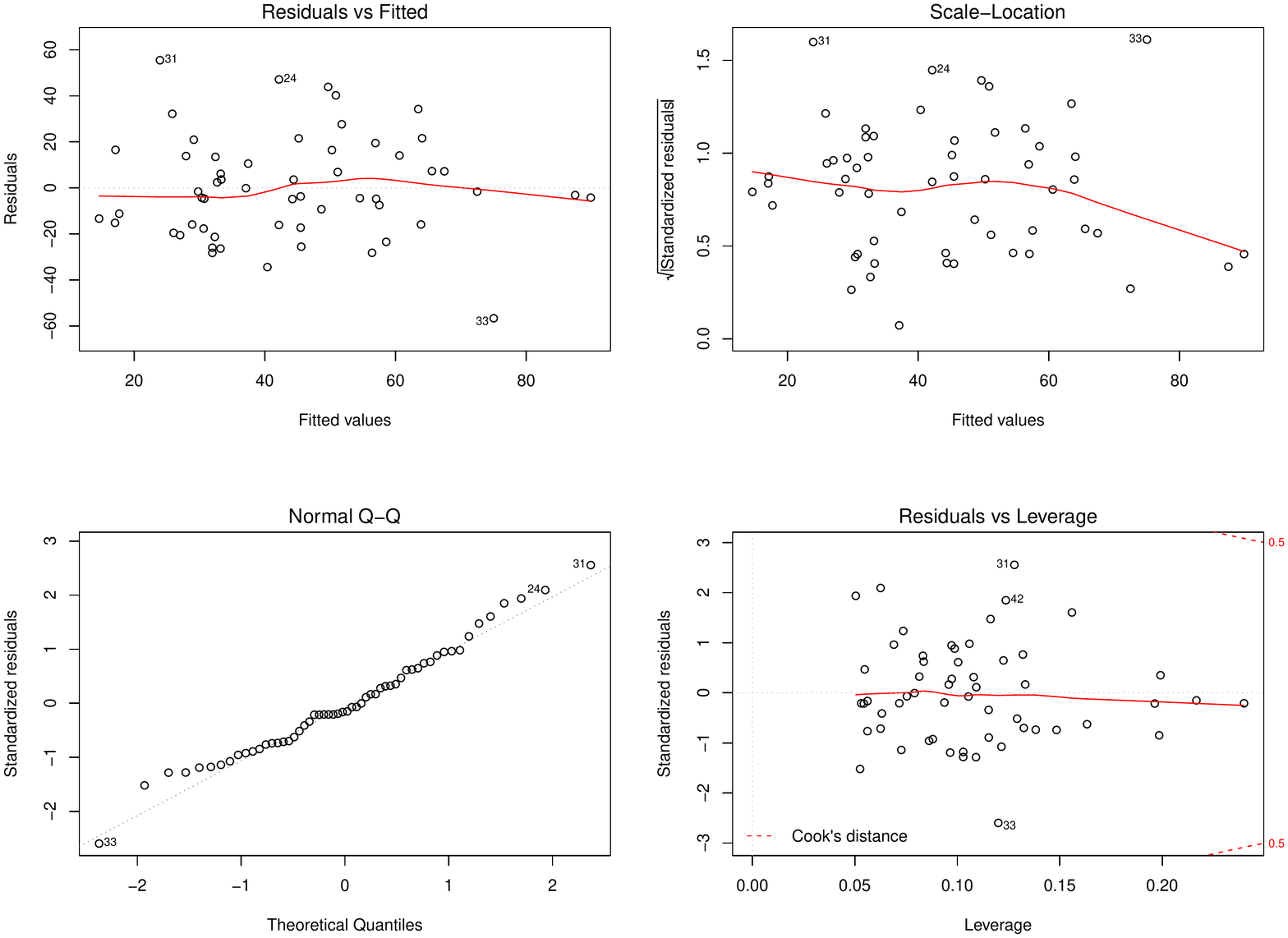}

\subsubsection{Phase 2 Efficacy: BMI}
\begin{verbatim}
Residuals:
    Min      1Q  Median      3Q     Max 
-29.599  -3.581   1.045   5.294  18.453 

Coefficients:
            Estimate Std. Error t value Pr(>|t|)    
(Intercept) 50.73675    1.73837  29.186   <2e-16 ***
TD          -5.26487    3.05415  -1.724   0.0909 .  
EXACER      -2.13863    2.86341  -0.747   0.4586    
BMI          0.92768    0.05073  18.286   <2e-16 ***
CLOZ        -1.11087    1.35825  -0.818   0.4173    
BMI:CLOZ    -0.05915    0.05248  -1.127   0.2650    
---
Signif. codes:  0 ‘***’ 0.001 ‘**’ 0.01 ‘*’ 0.05 ‘.’ 0.1 ‘ ’ 1 

Residual standard error: 9.611 on 50 degrees of freedom
Multiple R-squared: 0.8846,	Adjusted R-squared: 0.8731 
F-statistic: 76.67 on 5 and 50 DF,  p-value: < 2.2e-16 
\end{verbatim}
\includegraphics[width=\diagfigwidth\textwidth]{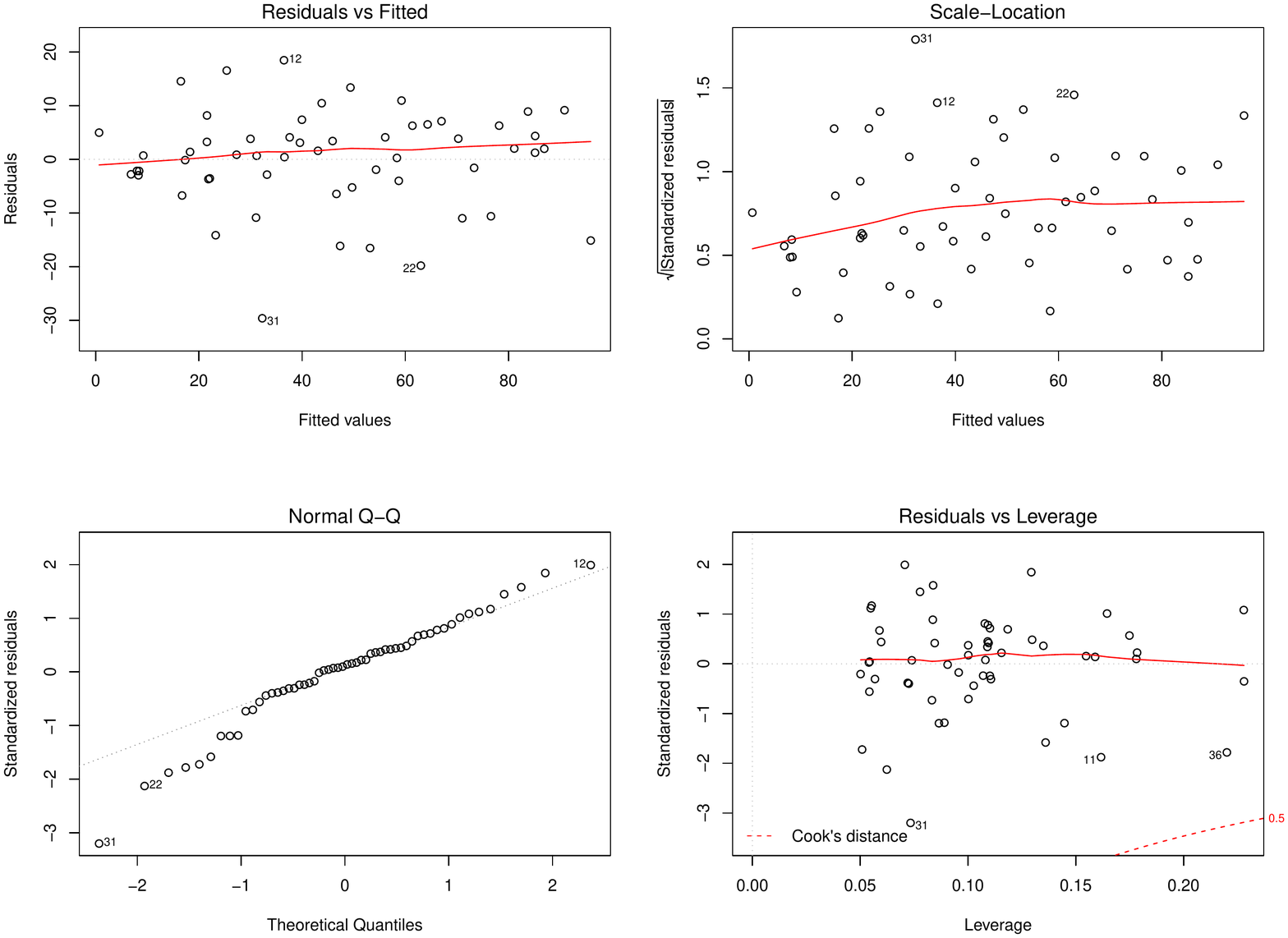}

\subsubsection{Phase 1 PANSS - For a particular feasible Phase 2  decision rule} 
\begin{verbatim}
Residuals:
    Min      1Q  Median      3Q     Max 
-71.066 -13.755   0.302  15.982  66.103 

Coefficients:
            Estimate Std. Error t value Pr(>|t|)    
(Intercept) 57.35755    1.03817  55.249  < 2e-16 ***
TD          -5.88398    2.04147  -2.882  0.00404 ** 
EXACER       1.12338    1.64705   0.682  0.49537    
PANSS        0.53324    0.03182  16.757  < 2e-16 ***
PERP        -2.66918    0.95054  -2.808  0.00508 ** 
PANSS:PERP   0.07784    0.03173   2.453  0.01434 *  
---
Signif. codes:  0 ‘***’ 0.001 ‘**’ 0.01 ‘*’ 0.05 ‘.’ 0.1 ‘ ’ 1 

Residual standard error: 22.53 on 969 degrees of freedom
Multiple R-squared: 0.2854,	Adjusted R-squared: 0.2817 
F-statistic: 77.38 on 5 and 969 DF,  p-value: < 2.2e-16 
\end{verbatim}
\includegraphics[width=\diagfigwidth\textwidth]{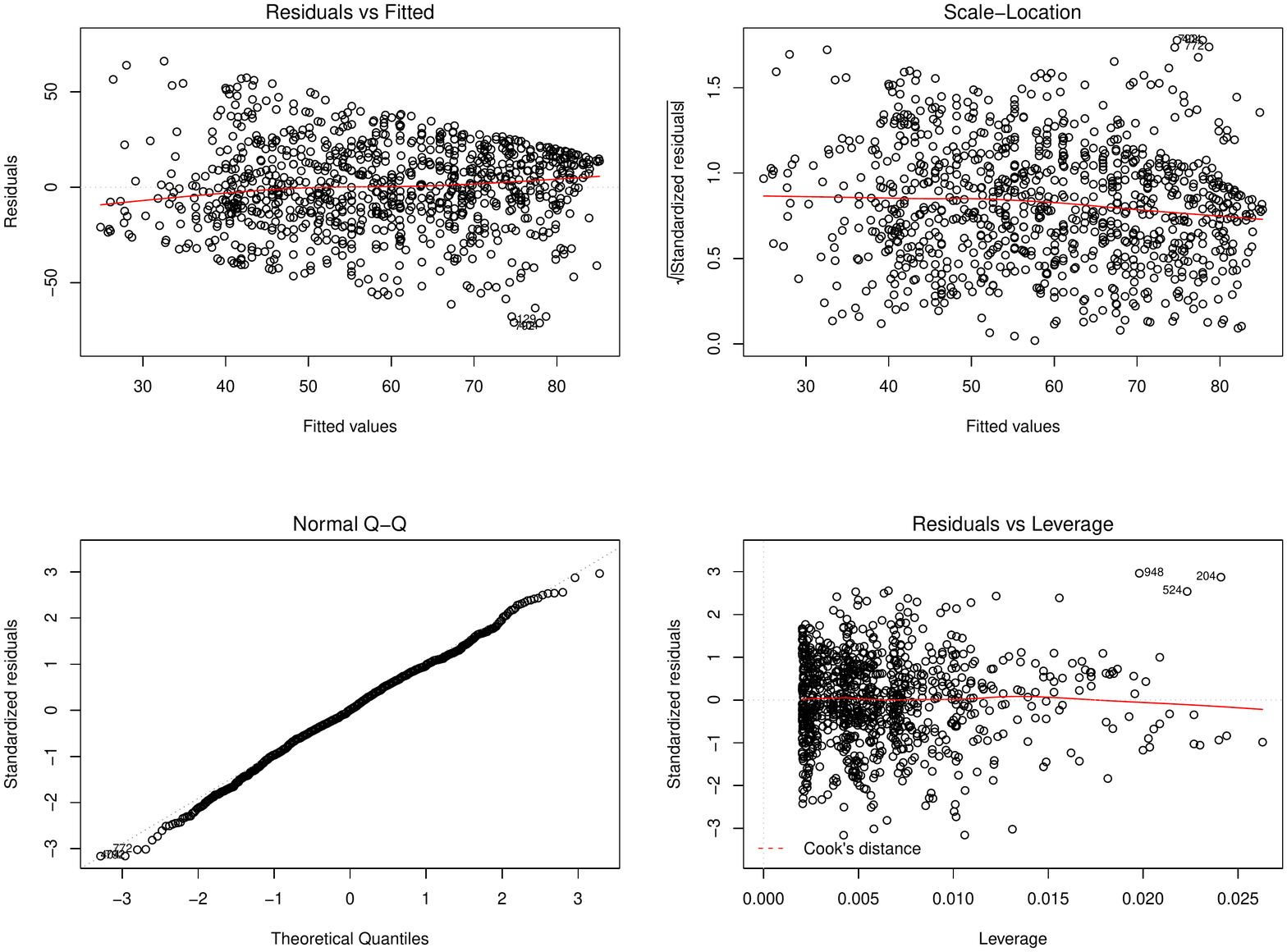}

\subsubsection{Phase 1 BMI - For a particular feasible Phase 2 decision rule}
\begin{verbatim}
Residuals:
    Min      1Q  Median      3Q     Max 
-59.767  -4.776  -0.307   5.721  48.318 

Coefficients:
            Estimate Std. Error t value Pr(>|t|)    
(Intercept) 49.06219    0.50837  96.509  < 2e-16 ***
TD           1.20040    1.00689   1.192 0.233479    
EXACER      -2.78115    0.81747  -3.402 0.000696 ***
BMI          0.90136    0.01632  55.240  < 2e-16 ***
PERP         1.82645    0.46597   3.920 9.49e-05 ***
BMI:PERP    -0.02502    0.01626  -1.539 0.124167    
---
Signif. codes:  0 ‘***’ 0.001 ‘**’ 0.01 ‘*’ 0.05 ‘.’ 0.1 ‘ ’ 1 

Residual standard error: 11.13 on 969 degrees of freedom
Multiple R-squared: 0.8451,	Adjusted R-squared: 0.8443 
F-statistic:  1057 on 5 and 969 DF,  p-value: < 2.2e-16 
\end{verbatim}
\includegraphics[width=\diagfigwidth\textwidth]{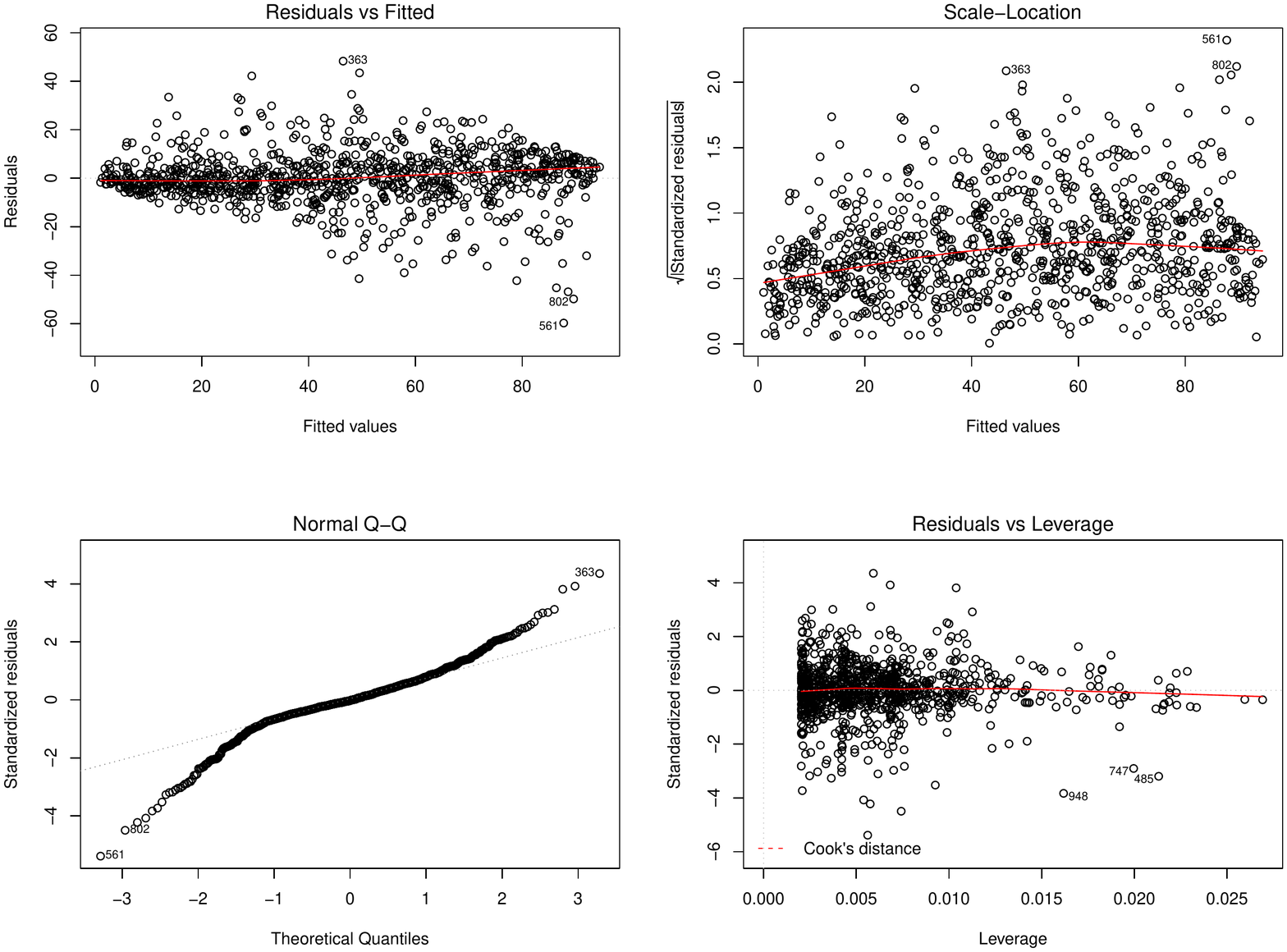}

\end{document}